\def\rfr#1{eq. (\ref{#1})}

\def\virg#1{``#1''}

\def\eqi{\begin{equation}}
\def\eqf{\end{equation}}
\def\eqia{\begin{eqnarray}}
\def\eqfa{\end{eqnarray}}
\def\rp#1#2{{#1\over#2}} \def\lb#1{\label{#1}}

\def\bds#1{\boldsymbol{#1}}

\documentclass{ws-ijmpd} 
\usepackage{latexsym}
\usepackage{graphicx,epsfig}

\RequirePackage{color}
\begin{document}

\markboth{L. Iorio}
{Effects of Standard and Modified Gravity on Interplanetary Ranges}

%
%

\title{EFFECTS OF STANDARD AND MODIFIED GRAVITY ON INTERPLANETARY RANGES }

\author{LORENZO IORIO}

\address{Ministero dell'Istruzione, dell'Universit\`{a} e della Ricerca (M.I.U.R.), Fellow of the Royal Astronomical Society (F.R.A.S.). Permanent address: Viale Unit\`{a} di Italia 68\\
Bari, (BA) 70125,
Italy\\
lorenzo.iorio@libero.it}

\maketitle

\begin{history}
\received{22 May 2010}
\revised{25 October 2010}
\accepted{5 November 2010}
\comby{J. Pullin}
\end{history}

\begin{abstract}
We numerically investigate the impact on the two-body range  of several Newtonian and non-Newtonian dynamical effects for some Earth-planet \textcolor{black}{(Mercury, Venus, Mars, Jupiter, Saturn)} pairs in view
of the expected cm-level accuracy in \textcolor{black}{some} future planned or proposed interplanetary ranging operations. The general relativistic gravitomagnetic Lense-Thirring effect should be modeled and solved-for in future accurate ranging tests of Newtonian and post-Newtonian gravity because it falls within their measurability domain. It could a-priori \virg{imprint} the determination of some of the target parameters of the tests considered. Moreover, the ring of the minor asteroids, Ceres, Pallas, Vesta \textcolor{black}{(and also many other asteroids if Mars is considered)} and the Trans-Neptunian Objects (TNOs) act as sources of nonnegligible systematic uncertainty on the larger gravitoelectric post-Newtonian signals from which it is intended to determine the parameters $\gamma$ and $\beta$ of the Parameterized Post\textcolor{black}{-}Newtonian (PPN) formalism with very high precision (\textcolor{black}{several} orders of magnitude better than the current $10^{-4}-10^{-5}$ levels). Also other putative, nonconventional gravitational effects like a violation of the Strong Equivalence Principle (SEP), a secular variation of the Newtonian constant of gravitation $G$, and the Pioneer anomaly are considered. The presence of a hypothetical, distant planetary-sized body X could be detectable with future high-accuracy planetary ranging. Our analysis can, in principle,  be extended
also to future interplanetary ranging scenarios in which one or more spacecraft in heliocentric orbits are involved. \textcolor{black}{The impact of fitting the initial conditions, and of the noise in the observations, on the actual detectability of the dynamical signatures investigated, which may be partly absorbed in the estimation process, should be quantitatively addressed in further studies. }
\end{abstract}


\section{Introduction}
Recent years have seen increasing efforts towards the implementation of the Planetary Laser Ranging (PLR) technique accurate to cm-level\cite{ILR1,ILR2,ILR,Degn,ILR3,ILR4,MOLA,recent}. It would allow to reach major improvements in three related fields: solar system dynamics, tests of general relativity and alternative theories of gravity, and physical properties of the target planet itself. In principle, any solar system body endowed  with a solid surface and a transparent atmosphere would be a suitable platform for a PLR system, but some targets are more accessible than others. Major efforts have been practically devoted so far to Mercury\cite{ILR1} and Mars\cite{ILR2,ILR3}, although simulations reaching 93 a.u. or more have been undertaken as well\cite{Degn,MOLA}.
In 2005 two interplanetary laser transponder experiments were successfully
demonstrated by the Goddard Geophysical Astronomical Observatory (GGAO). The first utilized the nonoptimized Mercury Laser Altimeter
(MLA) on the Messenger spacecraft\cite{ILR1,ILR}, obtaining a formal error in the laser range solution of 0.2 m, or one part in $10^{11}$. The second utilized the Mars Orbiting Laser
Altimeter (MOLA) on the Mars Global Surveyor spacecraft\cite{Abshire,ILR}. A precise measure of the Earth-Mars distance, measured between their centers of mass and taken over an extended period (five years or more), would support, among other things, a better determination of several parameters of the solar system. Sensitivity analyses point towards measurement uncertainties between\cite{ILR2} 1 mm and 100 mm.
\textcolor{black}{The perspectives in measuring the distance between an Earth-based station and an active laser transponder on the Martian moon Phobos capable of reaching mm-level range resolution have  recently been investigated in Ref.~\refcite{turyllo}. The authors of Ref.~\refcite{getemme}  envisage the possibility of using also Deimos, in addition to Phobos, as a target for an in-situ lander in the framework of the proposed Gravity Experiment with TimE Metrology on Martian satEllites (GETEMME) mission\footnote{See on the WEB http://meetingorganizer.copernicus.org/EPSC2010/EPSC2010-60.pdf.}. Its goal is to use laser to measure intermartian distances with an accuracy of a few tenth of mm, and its nominal duration should be $3-5$ yr. See also Ref.~\refcite{Ioriomars} for an earlier, preliminary study on the possibility of using both artificial and natural satellites to test general relativity in the martian system. }
Concerning Mercury, a recent analysis on the future BepiColombo\footnote{It is an ESA mission, including two spacecraft, one of which provided by Japan, to be put into orbit around Mercury. The launch is scheduled for 2014. The construction of the instruments is currently ongoing.} mission, aimed to accurately determining, among other things, several key parameters of post-Newtonian gravity and the solar quadrupole moment from Earth-Mercury distance data collected with  a multi-frequency radio link\cite{Milani0,Milani}, points toward a maximum uncertainty of $4.5-10$ cm in determining the Earth-Mercury range over a multi-year time span\cite{Milani0,Ashby,Milani} (1-8 yr). A proposed spacecraft-based mission aimed to accurately measure also  the \textcolor{black}{general relativistic} gravitomagnetic field of the Sun and its  \textcolor{black}{adimensional quadrupole mass moment} $J_2$  along with other PPN parameters like $\gamma$ and $\beta$ by means of interplanetary ranging is the Astrodynamical Space Test of
Relativity using Optical Devices\footnote{Its cheaper version ASTROD I makes use of one
spacecraft in a Venus-gravity-assisted solar orbit, ranging optically with ground
stations\cite{astrod1}.} (ASTROD)\cite{astrod}.
Another space-based missions proposed to accurately test several aspects of  the gravitational interaction via interplanetary laser ranging are the Laser Astrometric Test of Relativity (LATOR)\cite{lator}, \textcolor{black}{ and the interplanetary range to Phobos\cite{turyllo} one goal of which is a measurement of $\gamma$ with an accuracy of $10^{-7}$. }
For a review of the motivations for accurately determining the parameters of post-Newtonian gravity, in particular $\beta$ and $\gamma$, see, e.g., Ref.~\refcite{Tur1,Tur2} and references therein.

In this paper we study the effects that several Newtonian and non-Newtonian dynamical features of motion have on the two-body range for the Earth and some planets of the solar system  for which accurate ranging to spacecraft exists or is planned in future.
Our goal is  to inspect the potential aliasing posed by other \textcolor{black}{competing} dynamical forces acting as source of systematic uncertainty. Indeed, it must be recalled that in the range observables actually used in testing post-Newtonian gravity there is also a part due to the Earth-planet orbital motions in addition to the  purely post-Newtonian Shapiro delay connected with the propagation of electromagnetic waves. Thus, reaching unprecedented accuracy in only measuring the latter effect is useless if the accuracy of the orbital signal is not at a comparable level.
\textcolor{black}{ On the other hand, it should be remarked that we do not aim to quantitatively assess the actual measurability of the dynamical effects investigated. It is a different and important task which would deserve a dedicated work. Indeed, a fit of the initial conditions to the real observations would be needed in order to realistically evaluate the level of removal of the effects of interest from the signatures. It is a nontrivial task which is beyond the scopes of the present analysis which could be fruitfully used to single out the most relevant dynamical signals and focus future efforts on them. Anyway, we will use in the following a heuristic rule-of-thumb. }
The paper is organized as follows. In Section \ref{method} we outline the strategy followed and mention the Newtonian and non-Newtonian effects investigated. In Section \ref{mercury}, Section \ref{venus}, Section \ref{mars}, \textcolor{black}{ Section \ref{jupiter}, } and Section \ref{saturn} we deal with the ranges of Mercury, Venus, Mars, \textcolor{black}{Jupiter} and Saturn, respectively. Section \ref{conclu} is devoted to the conclusions.
\section{Method}\lb{method}
In order to numerically obtain the effect of a given gravitational acceleration, considered as a relatively small perturbation P of the Newtonian Sun's monopole, on the range $|\vec{\rho}|$ between the Earth-Moon Barycenter (EMB) and a planet  we used MATHEMATICA to simultaneously integrate with the Runge-Kutta method the equations of motion in Cartesian coordinates of EMB and the planet considered with and without the perturbation P investigated by using the same set of initial conditions. We adopted the ICRF/J2000.0 reference frame, with the ecliptic and mean equinox of the reference epoch, i.e. J2000.0, centered at the Solar System Barycenter (SSB); the initial conditions at the epoch J2000.0 were retrieved with the HORIZONS WEB interface by JPL, NASA. The temporal interval of the numerical integration for \textcolor{black}{Mercury and Venus} has been taken equal to $\Delta t=2$ yr \textcolor{black}{in view of the fact that} the typical operational time spans \textcolor{black}{envisaged} for future PLR technique are similar. \textcolor{black}{For Mars and Saturn, for which ranging to spacecraft is currently ongoing, we adopted $\Delta t=5$ yr.} \textcolor{black}{Also for Jupiter we used $\Delta t=5$ yr.} The basic model adopted consists of the barycentric equations of motion of the Sun, the eight planets, the Moon, Ceres, Pallas, Vesta, Pluto and Eris, to be simultaneously integrated; the forces acting on them  include the mutual Newtonian $N-$body interactions, the perturbation due to the solar quadrupolar mass moment $J_2$, the effect of two rings modeling the actions of the minor asteroids and of the Trans-Neptunian Objects (TNOs), and the general relativistic gravitoelectric Schwarzschild and gravitomagnetic Lense-Thirring fields of the Sun. As additional perturbations, we modeled the action of a distant, planet-like body kept fixed in a given spatial position (planet X), secular rate $\dot G/G$  of the Newtonian gravitational constant $G$, a violation of the Strong Equivalence Principle (SEP) by means of the Nordtvedt parameter $\eta$, a constant and uniform acceleration radially directed towards the Sun, \textcolor{black}{and acting on Uranus, Neptune, Pluto (and Eris)}, to account for the Pioneer anomaly.

In order to \textcolor{black}{preliminarily give an idea of} the potential measurability of the effects considered, the computed differences $\Delta |\vec{\rho}|\doteq |\vec{\rho}_{\rm P}|-|\vec{\rho}_{\rm R}|$, where R refers to a reference orbit which does not contain the perturbation P \textcolor{black}{of interest}, were subsequently compared to the available time series of the range residuals for the inner planets and Saturn which set the present-day accuracy level in ranging to planets\cite{DE421,Pit}. When the possibility that a given, unmodeled dynamical effect may show or not its signature in the range residuals it must be considered that the magnitude of such an effect should roughly be one order of magnitude larger than the range residuals accuracy. This to avoid the risk that it may be absorbed and partially or totally removed from the signature in the process of estimation of the initial conditions and of the other numerous solve-for parameters in the real data reduction. \textcolor{black}{ As outlined before, it is just a sort of rule of thumb; fully quantitative, realistic analyses, outside the scopes of the present paper, should require the actual fitting of the initial conditions to the observations.}

Depending on the dynamical effect one is interested in, some of the perturbations examined here are to be considered as sources of noise inducing systematic bias on the target signal. For example, if the goal of the analysis is, say, the Lense-Thirring effect, then the range perturbation due to the TNOs is clearly a source of potential systematic error which has to be evaluated. Thus, our plots are useful to assess the level of aliasing of several potential sources of aliasing for some non-Newtonian effects and the
correlations that may occur in estimating them. Dynamical effects which are viewed as noise in a given context can also be regarded as main targets in another one; see, e.g., the proposed  determination of asteroid masses through the ASTROD mission\cite{sassi}.
\section{Earth-Mercury range}\lb{mercury}
At present, the 1-way range residuals of Mercury from radar-ranging span 30 yr (1967-1997) and are at a few km-level (Figure B-2 of Ref.~\refcite{DE421}); the same holds for the 1-way Mercury radar closure residuals covering  8 yr (1989-1997, Figure B-3 a) of Ref.~\refcite{DE421}). There are also a pair of Mariner 10 range residuals in the 70s at Mercury at 0.2 km level (Figure B-3 b) of Ref.~\refcite{DE421}). Ranging to BepiColombo should be accurate to\cite{Ashby,Milani} $4.5-10$ cm over a few years.
\subsection{The Schwarzschild field of the Sun}
In Figure \ref{EMB_Mercury_Schwa} we plot the effect of  the gravitoelectric Schwarzschild field of the Sun on the Earth-Mercury range.
We modeled its acceleration as\cite{IERS}
\eqi\bds a_{\rm SS} = \rp{GM}{c^2 r^3}\left\{\left[2(\beta+\gamma)\rp{GM}{r}-\gamma \bds v\bds\cdot\bds v\right]\bds r+2(1+\gamma)\left(\bds r\bds\cdot\bds v\right)\bds v\right\},\eqf
where we inserted the PPN parameters $\beta$ and $\gamma$: they are equal to 1 in general relativity and we used such values.
\begin{figure}[pb]
\centerline{\psfig{file=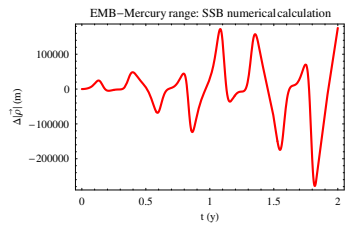
 ,width=4.7cm}}\vspace*{8pt}\caption{Difference $\Delta |\vec{\rho}|\doteq |\vec{\rho}_{\rm P}|-|\vec{\rho}_{\rm R}|$ in the numerically integrated EMB-Mercury ranges with and without the perturbation due to the Sun's Schwarzschild  field over $\Delta t=2$ yr. The same initial conditions (J2000.0) have been used for both the integrations. The state vectors at the reference epoch have been retrieved from the NASA JPL Horizons system. The integrations have been performed in the  ICRF/J2000.0 reference frame, with the ecliptic and mean equinox of the reference epoch, centered at the Solar System Barycenter (SSB). }\lb{EMB_Mercury_Schwa}
\end{figure}
  Figure \ref{EMB_Mercury_Schwa} can be compared with Figure 1 of Ref.~\refcite{Milani}, obtained for unspecified initial conditions\footnote{It also includes the Shapiro delay contribution.}: they are quite similar. The maximum variation of the signal is of the order of $4\times 10^5$ m, corresponding to a measurement accuracy of about $2.5\times 10^{-7}$. The expected realistic accuracy in determining $\beta$ and $\gamma$ is $2\times 10^{-6}$ in BepiColombo\cite{Milani0}.
\subsection{The oblateness of the Sun}
Figure \ref{EMB_Mercury_J2} shows the nominal effect of the Sun's quadrupolar mass moment  on the Mercury range for $J_2=2\times 10^{-7}$.
Its action has been modeled as\cite{VRBIK}
\eqi \bds a_{J_2}=-\rp{3J_2 R^2 GM}{2r^4}\left\{\left[1-5\left(\bds{\hat r}\bds\cdot\bds k\right)^2\right]\bds{\hat r}+2\left(\bds{\hat r}\bds\cdot\bds k\right)\bds k\right\}, \lb{U2}\eqf
where $R$ is the Sun's mean equatorial radius and $\bds k$ is the unit vector of the $z$ axis directed along the body's rotation axis. Since \rfr{U2} holds in a frame with its $\{xy\}$ plane coinciding with the body's equator, we rotated the the mean ecliptic at the epoch to the Sun's equator which is inclined to it by the Carrington angle\cite{Carri} $i=7.15$ deg.
\begin{figure}[pb]
\centerline{\psfig{file=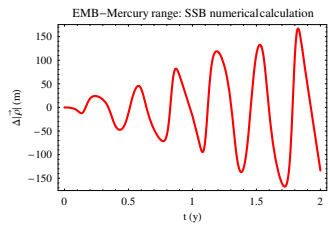
 ,width=4.7cm}}\vspace*{8pt}\caption{Difference $\Delta |\vec{\rho}|\doteq |\vec{\rho}_{\rm P}|-|\vec{\rho}_{\rm R}|$ in the numerically integrated EMB-Mercury ranges with and without the nominal perturbation due to the Sun's quadrupole mass moment $J_2=2.0\times 10^{-7}$ over $\Delta t=2$ yr. The same initial conditions (J2000.0) have been used for both the integrations. The state vectors at the reference epoch have been retrieved from the NASA JPL Horizons system. The integrations have been performed in the  ICRF/J2000.0 reference frame, with the mean equinox of the reference epoch and the reference $\{xy\}$ plane rotated from the mean ecliptic of the epoch to the Sun's equator, centered at the Solar System Barycenter (SSB). }\lb{EMB_Mercury_J2}
\end{figure}
 The signal of Figure \ref{EMB_Mercury_J2} has a maximum span of 300 m, corresponding to an accuracy measurement of $3\times 10^{-4}$. A measure of the solar $J_2$ accurate to  $10^{-2}$  is one of the goals of BepiColombo\cite{Milani0}; knowing precisely $J_2$ would yield important insights on the internal rotation of the Sun. At present, it is known with an uncertainty of about\cite{Fienga} $10\%$. The solar quadrupole mass moment may play the role of source of systematic bias with respect to, e.g., some non-Newtonian dynamical effects. Concerning the gravitoelectric signal previously analyzed, the mismodeled $J_2$ signature would impact it a $7.5\times 10^{-5}$ level. It is important to note that the patterns of the two signals are rather different.  Conversely, as we will see, the determination of $J_2$ at the desired level of accuracy may be affected by other unmodeled/mismdeled dynamical effects acting as systematic sources of aliasing on it.
\subsection{The Lense-Thirring effect of the Sun}
Figure \ref{EMB_Mercury_LT} depicts the range perturbation due to the Sun's Lense-Thirring effect, neither considered so far  in the dynamical force models of the planetary ephemerides nor in the BepiColombo analyses. It is a general relativistic feature of motion induced by the rotation of the Sun which acts upon a test particle moving with velocity $\bds v$ with a noncentral acceleration\cite{IERS}
\eqi \bds a_{\rm LT}=\rp{(1+\gamma)G}{c^2 r^3}\left[\rp{3}{r^2}\left(\bds r\bds\times\bds v\right)\left(\bds r\bds\cdot\bds S\right)+\left(\bds v\bds\times\bds S\right)\right],\lb{accleti}\eqf where $\bds S$ is the Sun's proper angular momentum. According to helioseismology\cite{Pij1,Pij2}, its magnitude is $S=(190.0\pm 1.5)\times 10^{39}$ kg m$^2$ s$^{-1}$.
\begin{figure}[pb]
\centerline{\psfig{file=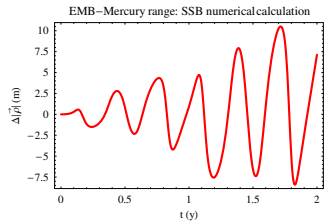
 ,width=4.7cm}}\vspace*{8pt}\caption{Difference $\Delta |\vec{\rho}|\doteq |\vec{\rho}_{\rm P}|-|\vec{\rho}_{\rm R}|$ in the numerically integrated EMB-Mercury ranges with and without the perturbation due to the Sun's Lense-Thirring field over $\Delta t=2$ yr. The same initial conditions (J2000.0) have been used for both the integrations. The state vectors at the reference epoch have been retrieved from the NASA JPL Horizons system. The integrations have been performed in the  ICRF/J2000.0 reference frame, with the mean equinox of the reference epoch and the reference $\{xy\}$ plane rotated from the mean ecliptic of the epoch to the Sun's equator, centered at the Solar System Barycenter (SSB). }\lb{EMB_Mercury_LT}
\end{figure}
Note that such a value does not come from planetary orbital dynamics, so that there is no risk of a-priori \virg{imprinting} of  general relativity itself on range tests of the solar Lense-Thirring effect which could, thus, be regarded as genuine and unbiased. Also in this case we rotated the reference frame to the mean ecliptic at the epoch to the Sun's equator by the Carrington angle because \rfr{accleti} holds in a frame with its $z$ axis aligned with $\bds S$. The peak-to-peak amplitude of the Lense-Thirring signal  is up to $17.5$ m over 2 yr, which, if on the one hand is unmeasurable from currently available radar-ranging to Mercury, on the other hand corresponds to a potential relative accuracy in measuring it with BepiColombo of $2-5\times 10^{-3}$; this clearly
shows that the solar gravitomagnetic  field should be taken into account in future analyses and data processing. Otherwise, it would alias the recovery of other effects. For example, it may affect the determination of $J_2$ at $12\%$ level. On the other hand, in order to allow for a determination of the Lense-Thirring effect,
the Sun's quadrupole mass moment should be known with an accuracy better than the present-day one by at least one order of magnitude; this is just one of the goals of BepiColombo.
Moreover, since the Lense-Thirring effect depends on\footnote{The multiplicative factor 2 in front of \rfr{accleti} comes from\cite{IERS} $1+\gamma$. } $\gamma$, neglecting it may alias the determination of $\gamma$ through the larger gravitoelectric  signal at $4\times 10^{-5}$ level.
\subsection{The ring of the minor asteroids and Ceres, Pallas and Vesta}
In Figure \ref{EMB_Mercury_astring} we depict one potential source of systematic bias, i.e. the action of the ring of minor asteroids\cite{Fienga}.
We modeled it \textcolor{black}{following Ref.~\refcite{Fienga08}}. For those planets for which $r>R_{\rm ring}$, by posing $\alpha\doteq R_{\rm ring}/r$, we obtained
\eqi \bds a_{\rm inner\ ring}\simeq -\rp{Gm_{\rm ring}}{2r^3}\left(2+\rp{3}{2}\alpha^2+\rp{45}{32}\alpha^4\right)\bds r, \eqf
from
\eqi \bds a_{\rm inner\ ring}=\rp{Gm_{\rm ring}}{2r^3}\left[\alpha b^{(1)}_{\rp{3}{2}}(\alpha)-b^{(0)}_{\rp{3}{2}}(\alpha)\right]\bds r.\eqf
For $r<R_{\rm ring}$, by posing $\alpha\doteq r/R_{\rm ring}$, we obtained
\eqi \bds a_{\rm outer\ ring}\simeq \rp{Gm_{\rm ring}}{2rR_{\rm ring}^2}\left(\alpha+\rp{9}{8}\alpha^3+\rp{75}{64}\alpha^5\right)\bds r, \eqf
from
\eqi \bds a_{\rm outer\ ring}=\rp{Gm_{\rm ring}}{2rR_{\rm ring}^2}\left[b^{(1)}_{\rp{3}{2}}(\alpha)-\alpha b^{(0)}_{\rp{3}{2}}(\alpha)\right]\bds r.\eqf
Recall that the Laplace coefficients are defined as
\eqi b_s^{(j)}(\alpha)\doteq\rp{1}{\pi}\int_0^{2\pi}\rp{\cos j\psi d\psi}{\left(1-2\alpha\cos\psi+\alpha^2\right)^s},\eqf where $s$ is a half-integer; a useful approximate expression in terms of a series is\cite{Murr}
\eqi b_s^{(j)}\simeq\rp{s(s+1)...(s+j-1)}{1\cdot3\cdot\cdot\cdot j}\alpha^j\left[1+\rp{s(s+j)}{(1+j)}\alpha^2+\rp{s(s+1)(s+j)(s+j+1)}{1\cdot 2(j+1)(j+2)}\alpha^4\right].\eqf
\begin{figure}[pb]
\centerline{\psfig{file=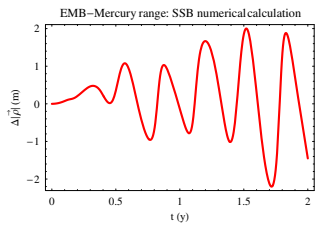
 ,width=4.7cm}}\vspace*{8pt}\caption{Difference $\Delta |\vec{\rho}|\doteq |\vec{\rho}_{\rm P}|-|\vec{\rho}_{\rm R}|$ in the numerically integrated EMB-Mercury ranges with and without the nominal perturbation due to the ring of minor asteroids with\protect\cite{Fienga} $m_{\rm ring}=1\times 10^{-10}$M$_{\odot}$  and $R_{\rm ring}=3.14$ a.u. over $\Delta t=2$ yr. The same initial conditions (J2000.0) have been used for both the integrations. The state vectors at the reference epoch have been retrieved from the NASA JPL Horizons system. The integrations have been performed in the  ICRF/J2000.0 reference frame, with the ecliptic and mean equinox of the reference epoch, centered at the Solar System Barycenter (SSB). }\lb{EMB_Mercury_astring}
\end{figure}
By assuming for the ring of the minor asteroids a nominal mass of\cite{Fienga} $m_{\rm ring}=1\times 10^{-10}$M$_{\odot}$  and a radius\cite{Fienga} $R_{\rm ring}=3.14$ au,
it would impact the Mercury range at 4 m level (peak-to-peak amplitude), which is, in fact, measurable. Its nominal bias on the Schwarzschild, $J_2$ and Lense-Thirring signals would be $1\times 10^{-5}, 1.3\times 10^{-2},2.3\times 10^{-1}$, respectively. Anyway, the present-day level of uncertainty in the mass of the ring is\cite{Fienga} $\delta m_{\rm ring}=0.3\times 10^{-10}$M$_{\odot}$. Thus, the impact of such a mismodeling would be, $3\times 10^{-6}, 4\times 10^{-3}, 7\times 10^{-2}$, respectively; it cannot be considered negligible.

The effect of  Ceres, Pallas and Vesta on the determination of some Newtonian and non-Newtonian parameters with BepiColombo has been preliminarily investigated in Ref.~\refcite{Ashby}. Here
in Figure \ref{EMB_Mercury_CePaVe}
we
show
the nominal perturbation on the Earth-Mercury range due to the combined actions of Ceres, Pallas and Vesta; the values for their masses have been retrieved from Ref.~\refcite{CePaVe}.
\begin{figure}[pb]
\centerline{\psfig{file=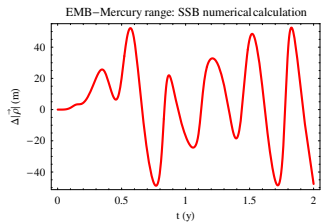
 ,width=4.7cm}}\vspace*{8pt}\caption{Difference $\Delta |\vec{\rho}|\doteq |\vec{\rho}_{\rm P}|-|\vec{\rho}_{\rm R}|$ in the numerically integrated EMB-Mercury ranges with and without the nominal perturbation due to\protect\cite{CePaVe}  Ceres, Pallas, Vesta  over $\Delta t=2$ yr. The same initial conditions (J2000.0) have been used for both the integrations. The state vectors at the reference epoch have been retrieved from the NASA JPL Horizons system. The integrations have been performed in the  ICRF/J2000.0 reference frame, with the ecliptic and mean equinox of the reference epoch, centered at the Solar System Barycenter (SSB). }\lb{EMB_Mercury_CePaVe}
\end{figure}
Its peak-to-peak amplitude amounts to 80 m; thus, their signature would be measurable at a $0.6-1\times 10^{-3}$ level. Anyway, the mismodeled solar quadrupole mass moment would bias their signal at $4\times 10^{-1}$ level. The Lense-Thirring effect, if unmodeled, would have an impact at $2.2\times 10^{-1}$ level.
The present-day relative uncertainties in their masses are $6\times 10^{-3}, 3\times 10^{-2}, 2\times 10^{-2}$ respectively\cite{CePaVe}. This implies a mismodeled signal with a peak-to-peak amplitude of 50 cm. It would impact the Schwarzschild, $J_2$ and Lense-Thirring range perturbations at $1\times 10^{-6},2\times 10^{-3},3\times 10^{-2}$ level, respectively.
\subsection{The Trans-Neptunian Objects}
The situation is different for another potential source of systematic uncertainty, i.e. the Trans-Neptunian Objects (TNOs). Figure  \ref{EMB_Mercury_tnoring}, obtained by modeling them as a ring with\cite{Pit} $m_{\rm ring}=5.26\times 10^{-8}$M$_{\odot}$  and\cite{Pit} $R_{\rm ring}=43$ au, shows that their maximum effect would amount to 80 cm. We used the same formulas as for the asteroid ring.
\begin{figure}[pb]
\centerline{\psfig{file=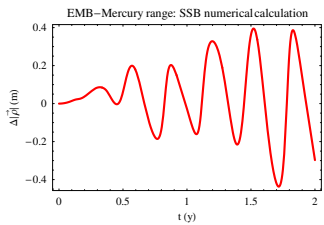
 ,width=4.7cm}}\vspace*{8pt}\caption{Difference $\Delta |\vec{\rho}|\doteq |\vec{\rho}_{\rm P}|-|\vec{\rho}_{\rm R}|$ in the numerically integrated EMB-Mercury ranges with and without the nominal perturbation due to the ring of Trans-Neptunian Objects with\protect\cite{Pit}  $m_{\rm ring}=5.26\times 10^{-8}$M$_{\odot}$ and $R_{\rm ring}=43$ a.u. over $\Delta t=2$ yr. The same initial conditions (J2000.0) have been used for both the integrations. The state vectors at the reference epoch have been retrieved from the NASA JPL Horizons system. The integrations have been performed in the  ICRF/J2000.0 reference frame, with the ecliptic and mean equinox of the reference epoch, centered at the Solar System Barycenter (SSB). }\lb{EMB_Mercury_tnoring}
\end{figure}
Such an effect, not taken into account so far, would be better measurable than that by the minor asteroids. This implies a bias of $2\times 10^{-6}$ on the Schwarzschild signal, $3\times 10^{-3}$ for $J_2$ and $4.5\times 10^{-2}$ for the Lense-Thirring effect. A major concern is that the mass of the TNOs is far from being accurately known, so that an uncertainty as large as $100\%$ should be applied.
\subsection{Violation of the Strong Equivalence Principle}
Let us, now, focus our attention to other nonstandard effects like a SEP violation, a variation of the Newtonian gravitational constant, the Pioneer anomaly and a putative planet X.

Concerning the SEP violation, we  modeled it in the acceleration $\bds a_i$ of a body $i$ as
\eqi \bds a^{\rm (SEP)}_{i}=\left[\rp{m^{(G)}_i}{m^{(I)}_i}\right]_{\rm SEP}\left(\sum_{j,i\neq j}\rp{Gm_{j}}{r^3_{ij}}\bds r_{ij}\right), i=1,2,...N.\eqf
In it $m^{(G)}_i$ and $m^{(I)}_i$ are the gravitational and inertial masses, respectively, of the body $i$. Their ratio is
\eqi \left[\rp{m^{(G)}_i}{m^{(I)}_i}\right]_{\rm SEP}=1+\eta\Omega_i\eqf
in which $\eta$ is the Nordtvedt dimensionless constant \cite{Nor1,Nor2} accounting for SEP violation\footnote{In terms of the PPN parameters $\beta$ and $\gamma$ it is $\eta=4\beta-\gamma-3$, so that $\eta=0$ in general relativity.}, and
\eqi \Omega_i\doteq\rp{E_i}{m_ic^2},\eqf
where $E_i$ is the  (negative) gravitational self-energy of the $i-$th body, and $m_ic^2$ is its total mass-energy.
For a spherical body of radius $R_i$ \cite{Tur1}, \eqi\Omega_i=-\rp{3}{5}\rp{Gm_i}{R_i c^2}.\eqf
Figure \ref{EMB_Mercury_eta} shows its effect on the Mercury range for $\eta=10^{-5}$; at present, the most accurate constraints on $\eta$ come from Lunar Laser Ranging (LLR) amounting to\cite{Will} $\eta=(4.0\pm 4.3)\times 10^{-4}$.
\begin{figure}[pb]
\centerline{\psfig{file=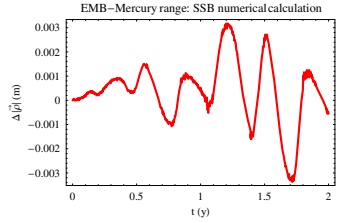
 ,width=4.7cm}}\vspace*{8pt}\caption{Difference $\Delta |\vec{\rho}|\doteq |\vec{\rho}_{\rm P}|-|\vec{\rho}_{\rm R}|$ in the numerically integrated EMB-Mercury ranges with and without the perturbation due to a violation of the Strong Equivalence Principle for $\eta=10^{-5}$ over $\Delta t=2$ yr. The same initial conditions (J2000.0) have been used for both the integrations. The state vectors at the reference epoch have been retrieved from the NASA JPL Horizons system. The integrations have been performed in the  ICRF/J2000.0 reference frame, with the ecliptic and mean equinox of the reference epoch, centered at the Solar System Barycenter (SSB). }\lb{EMB_Mercury_eta}
\end{figure}
 The peak-to-peak amplitude of the signature of Figure \ref{EMB_Mercury_eta} is 6 mm, which is practically impossible to detect. Moreover, the SEP signal would be totally swamped by other dynamical effects like the ones by the minor asteroid ring, even if modeled at the present-day level of accuracy, and the TNOs. Also the Lense-Thirring effect, modeled or not, would be of concern.
\subsection{Secular variation of the Newtonian constant of gravitation}
The case of a possible variation of the Newtonian gravitational constant is interesting because recently E.V. Pitjeva in Ref.~\refcite{Pit} preliminarily reported a secular variation for it $\dot G/G=(-5.9\pm 4.4)\times 10^{-14}$ yr$^{-1}$, statistically significant at $3-\sigma$ level. Other researchers get results statistically compatible with 0. Folkner in Ref.~\refcite{Fol} gets an upper bound of $2\times 10^{-13}$ yr$^{-1}$ from planetary ephemerides as well. Williams et al. in Ref.~\refcite{Will} obtain $\dot G/G=(4\pm 9)\times 10^{-13}$ yr$^{-1}$ from LLR.
We modeled a secular variation of $G$ in the equations of motion according to
\eqi \bds a^{ (\dot G)}_{i}=\sum_{j,i\neq j}\rp{G\left[1+\rp{\dot G}{G}(t-t_0)\right]m_{j}}{r^3_{ij}}\bds r_{ij}, i=1,2,...N.\eqf
Figure \ref{EMB_Mercury_Gdot} depicts its impact on the Mercury range.
\begin{figure}[pb]
\centerline{\psfig{file=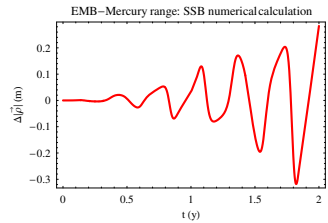
 ,width=4.7cm}}\vspace*{8pt}\caption{Difference $\Delta |\vec{\rho}|\doteq |\vec{\rho}_{\rm P}|-|\vec{\rho}_{\rm R}|$ in the numerically integrated EMB-Mercury ranges with and without the nominal perturbation due to a secular variation of $G$ as large as\protect\cite{Pit} $\dot G/G=-5.9\times 10^{-14}$ yr$^{-1}$  over $\Delta t=2$ yr. The same initial conditions (J2000.0) have been used for both the integrations. The state vectors at the reference epoch have been retrieved from the NASA JPL Horizons system. The integrations have been performed in the  ICRF/J2000.0 reference frame, with the ecliptic and mean equinox of the reference epoch, centered at the Solar System Barycenter (SSB). }\lb{EMB_Mercury_Gdot}
\end{figure}
Its maximum effect would be about 60 cm, which, in principle, should be measurable with BepiColombo at a $0.7-2\times 10^{-1}$ level of relative accuracy. If a secular decrease of $G$ will be confirmed as a genuine physical effect by further analyses of planetary data by independent teams of astronomers, it should be modeled in the BepiColombo data analysis because, otherwise, it would affect the Schwarzschild, $J_2$ and Lense-Thirring signatures at  $1\times 10^{-6},2\times 10^{-3},3\times 10^{-2}$ level, respectively. Anyway, the action of the Lense-Thirring effect, of Ceres, Pallas, Vesta and of TNOs, modeled or not, would likely bias the recovery of the putative $\dot G$ signal in a severe way; it must be recalled that the mismodeled signature due to the three large asteroids is 50 cm.
\subsection{The Pioneer Anomaly}
The Pioneer anomaly\cite{Pio}, which is a constant anomalous extra-acceleration approximately directed towards the Sun of magnitude $A_{\rm Pio}=8.74\times 10^{-10}$ m s$^{-2}$ detected in the telemetry of the Pioneer 10/11 spacecraft after they passed 20 au, may, in principle, impact the Earth-Mercury range as well in an indirect way through the altered action on them of the bodies directly affected by such a putative exotic force, i.e. Uranus, Neptune, Pluto and Eris. This would be another way of testing the hypothesis of a gravitational nature of the Pioneer anomaly in addition to directly looking at the outer planets\cite{IorGiu} which has given negative results\cite{StaPio,Fienga}.
In Figure \ref{EMB_Mercury_Pio} we plot its signature.
\begin{figure}[pb]
\centerline{\psfig{file=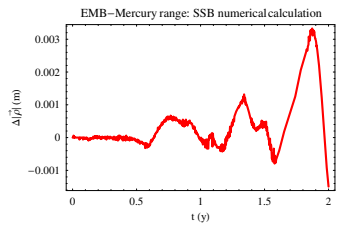
 ,width=4.7cm}}\vspace*{8pt}\caption{Difference $\Delta |\vec{\rho}|\doteq |\vec{\rho}_{\rm P}|-|\vec{\rho}_{\rm R}|$ in the numerically integrated EMB-Mercury ranges with and without the nominal perturbation due to a Pioneer-like constant and unform radial acceleration of\protect\cite{Pio} $A_{\rm Pio}=8.74\times 10^{-10}$ m s$^{-2}$ acting upon Uranus, Neptune, Pluto, Eris over $\Delta t=2$ yr. The same initial conditions (J2000.0) have been used for both the integrations. The state vectors at the reference epoch have been retrieved from the NASA JPL Horizons system. The integrations have been performed in the  ICRF/J2000.0 reference frame, with the ecliptic and mean equinox of the reference epoch, centered at the Solar System Barycenter (SSB). }\lb{EMB_Mercury_Pio}
\end{figure}
It would amount to 4 mm, which is too small to be realistically detected. Moreover, also the aliasing bias of the other effects previously considered would be crucial.
\subsection{Planet X}
Finally, let us consider the potential ability of BepiColombo of detecting the signature of a putative remote planet X. In
Figure \ref{EMB_Mercury_X} and
Figure \ref{EMB_Mercury_X2}
we depict the X's range signal for the
minimum and
maximum
value of its tidal parameter ${\mathcal{K}}_{\rm X}\doteq GM_{\rm X}/r_{\rm X}^3$
according to the anomalous perihelion precession of Saturn  analyzed in Ref.~\refcite{Iorio}.
\begin{figure}[pb]
\centerline{\psfig{file=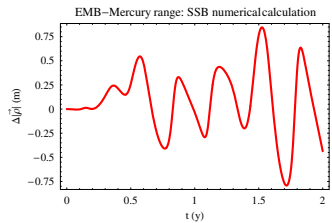
 ,width=4.7cm}}\vspace*{8pt}\caption{Difference $\Delta |\vec{\rho}|\doteq |\vec{\rho}_{\rm P}|-|\vec{\rho}_{\rm R}|$ in the numerically integrated EMB-Mercury ranges with and without the perturbation due to hypothetical remote planet X lying almost in the ecliptic with minimum tidal parameter\protect\cite{Iorio} ${\mathcal{K}}_{\rm X}=1.6\times 10^{-26}$
 s$^{-2}$  over $\Delta t=2$ yr. The same initial conditions (J2000.0) have been used for both the integrations. The state vectors at the reference epoch have been retrieved from the NASA JPL Horizons system. The integrations have been performed in the  ICRF/J2000.0 reference frame, with the ecliptic and mean equinox of the reference epoch, centered at the Solar System Barycenter (SSB). }\lb{EMB_Mercury_X}
\end{figure}
\begin{figure}[pb]
\centerline{\psfig{file=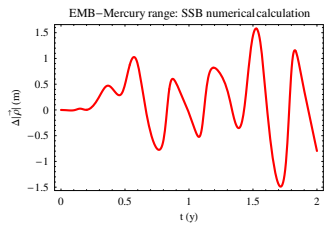
 ,width=4.7cm}}\vspace*{8pt}\caption{Difference $\Delta |\vec{\rho}|\doteq |\vec{\rho}_{\rm P}|-|\vec{\rho}_{\rm R}|$ in the numerically integrated EMB-Mercury ranges with and without the perturbation due to hypothetical remote planet X lying almost in the ecliptic with maximum tidal parameter\protect\cite{Iorio} ${\mathcal{K}}_{\rm X}=2.7\times 10^{-26}$
 s$^{-2}$  over $\Delta t=2$ yr. The same initial conditions (J2000.0) have been used for both the integrations. The state vectors at the reference epoch have been retrieved from the NASA JPL Horizons system. The integrations have been performed in the  ICRF/J2000.0 reference frame, with the ecliptic and mean equinox of the reference epoch, centered at the Solar System Barycenter (SSB). }\lb{EMB_Mercury_X2}
\end{figure}
Let us recall that recently Pitjeva in Ref.~\refcite{Pitjou} and Fienga et al. in Ref.~\refcite{Fienga}
independently determined  statistically significant extra-precessions of the perihelion of Saturn  from preliminary analysis of some years of radio-tracking data of the Cassini spacecraft. Their values are\footnote{E.V. Pitjeva, private communication, December 2008.} $\Delta\dot\varpi_{\rm Sat}=-6\pm 2$ mas cty$^{-1}$, cited in Ref.~\refcite{Fienga}, and\cite{Fienga} $\Delta\dot\varpi_{\rm Sat}=-10\pm 8$ mas cty$^{-1}$. Further data analyses of longer  Cassini data records are required to confirm or disproof the existence of such an anomaly as a genuine physical effect\footnote{Pitjeva in Ref.~\refcite{Pit} reports a new value which, instead, is statistically compatible with 0, i.e. $\Delta\dot\varpi_{\rm Sat}=-10\pm 15$ mas cty$^{-1}$}.
The maximum effect of X on the Mercury range would be as large as $1.5-3$ m, falling within the measurability domain of BepiColombo. It must be noted that a large part of such a signal may be largely confused with the action of the TNOs. The Lense-Thirring effect, if not modeled, would be another source of serious systematic error as well.

\section{Earth-Venus range}\lb{venus}
Although, at present, no tests of interplanetary ranging to Venus have been practically performed,  contrary to Mercury and Mars, we prefer to treat also its case not only for completeness but also because simulations of interplanetary transponder and laser communications experiments via dual station ranging to SLR satellites covering also Venus have been implemented\cite{Degn,MOLA}.
Currently available radar-ranging normal points to Venus cover about 33 yr, from 1962 to 1995. The range residuals are depicted in Figure B-6 of Ref.~\refcite{DE421}; after having been as large as 15 km in the first 10 yr, they drop below 5 km in the remaining. Figure B-4 shows the range residuals to Venus Express at Venus from 2006 to 2008; the are below the 10 m level.
\subsection{The Schwarzschild field of the Sun}
According to Figure \ref{EMB_Venus_Schwa}, the peak-to-peak amplitude of the general relativistic Schwarzschild effect is $1.2\times 10^5$ m over $\Delta t=2$ m.
\begin{figure}[pb]
\centerline{\psfig{file=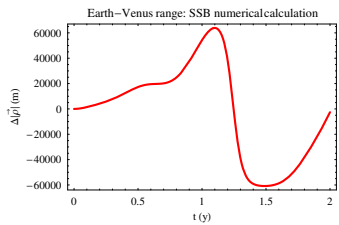
 ,width=4.7cm}}\vspace*{8pt}\caption{Difference $\Delta |\vec{\rho}|\doteq |\vec{\rho}_{\rm P}|-|\vec{\rho}_{\rm R}|$ in the numerically integrated EMB-Venus ranges with and without the perturbation due to the Sun's Schwarzschild  field over $\Delta t=2$ yr. The same initial conditions (J2000.0) have been used for both the integrations. The state vectors at the reference epoch have been retrieved from the NASA JPL Horizons system. The integrations have been performed in the  ICRF/J2000.0 reference frame, with the ecliptic and mean equinox of the reference epoch, centered at the Solar System Barycenter (SSB). }\lb{EMB_Venus_Schwa}
\end{figure}
A hypothetical, future laser ranging to, say, a suitably equipped target orbiting Venus accurate to 10 cm would allow to measure such a general relativistic signal with a relative accuracy of $8\times 10^{-7}$. In the following we will discuss the corrupting impact of some potential sources of systematic errors.
\subsection{The oblateness of the Sun}
The range perturbation due to the Sun's oblateness is depicted in Figure \ref{EMB_Venus_J2} for the nominal value $J_2=2\times 10^{-7}$. Also in this case a barycentric frame rotated to the Sun's equator has been adopted.
\begin{figure}[pb]
\centerline{\psfig{file=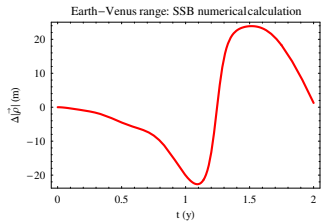
 ,width=4.7cm}}\vspace*{8pt}\caption{Difference $\Delta |\vec{\rho}|\doteq |\vec{\rho}_{\rm P}|-|\vec{\rho}_{\rm R}|$ in the numerically integrated EMB-Venus ranges with and without the nominal perturbation due to the Sun's quadrupole mass moment $J_2=2.0\times 10^{-7}$ over $\Delta t=2$ yr. The same initial conditions (J2000.0) have been used for both the integrations. The state vectors at the reference epoch have been retrieved from the NASA JPL Horizons system. The integrations have been performed in the  ICRF/J2000.0 reference frame, with the mean equinox of the reference epoch and the reference $\{xy\}$ plane rotated from the mean ecliptic of the epoch to the Sun's equator, centered at the Solar System Barycenter (SSB). }\lb{EMB_Venus_J2}
\end{figure}
The nominal maximum shift is 40 m, so that a measure accurate to $2.5\times 10^{-3}$ would be possible with a future 10 cm-level ranging technique. Viewed as a source of systematic uncertainty, the solar quadrupole mass moment would affect the Schwarzschild signal at $3\times 10^{-6}$ level by assuming the present-day uncertainty in it, i.e. $10\%$.
The temporal patterns of the two signals are quite different.
Note that the dynamical action of $J_2$ was modeled in producing the Venus Express residuals; thus, a mismodeled signal as large as just 4 m should have been left, in agreement with the range residuals.
\subsection{The Lense-Thirring effect of the Sun}
Figure \ref{EMB_Venus_LT} shows the Lense-Thirring perturbation of the Venus range, integrated in a frame aligned with the Sun's  equator.
\begin{figure}[pb]
\centerline{\psfig{file=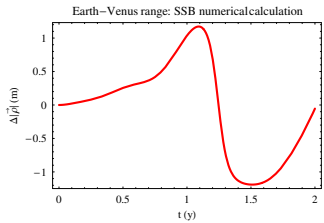
 ,width=4.7cm}}\vspace*{8pt}\caption{Difference $\Delta |\vec{\rho}|\doteq |\vec{\rho}_{\rm P}|-|\vec{\rho}_{\rm R}|$ in the numerically integrated EMB-Venus ranges with and without the perturbation due to the Sun's Lense-Thirring field over $\Delta t=2$ yr. The same initial conditions (J2000.0) have been used for both the integrations. The state vectors at the reference epoch have been retrieved from the NASA JPL Horizons system. The integrations have been performed in the  ICRF/J2000.0 reference frame, with the mean equinox of the reference epoch and the reference $\{xy\}$ plane rotated from the mean ecliptic of the epoch to the Sun's equator, centered at the Solar System Barycenter (SSB). }\lb{EMB_Venus_LT}
\end{figure}
The peak-to-peak amplitude  is 2 m, which would  be measurable with a future accurate cm-level ranging device with a relative accuracy of $2-5\times 10^{-2}$.  The Lense-Thirring signature is still too small to be detected nowadays with the current spacecraft ranging. If not modeled, the Sun's gravitomagnetic field would impact a determination of $J_2$ at a $5\%$ level, while the Schwarzschild signal would be biased by the Lense-Thirring one at a $3\times 10^{-5}$ level. It must be noted that the two relativistic signals exhibit very similar patterns. The present-day $10\%$ uncertainty in the Sun's oblateness  would yield  a mismodeled signal two times larger than the gravitomagnetic one. Anyway, their temporal signatures are different, so that it would be possible, in principle, to separate them.
\subsection{The ring of the minor asteroids and Ceres, Pallas and Vesta}
The impact of the ring of the minor asteroids on the Venus range is depicted in Figure \ref{EMB_Venus_astring}.
\begin{figure}[pb]
\centerline{\psfig{file=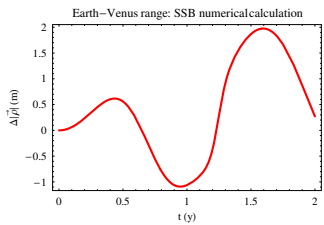
 ,width=4.7cm}}\vspace*{8pt}\caption{Difference $\Delta |\vec{\rho}|\doteq |\vec{\rho}_{\rm P}|-|\vec{\rho}_{\rm R}|$ in the numerically integrated EMB-Venus ranges with and without the nominal perturbation due to the ring of minor asteroids with\protect\cite{Fienga} $m_{\rm ring}=1\times 10^{-10}$M$_{\odot}$ and $R_{\rm ring}=3.14$ a.u. over $\Delta t=2$ yr. The same initial conditions (J2000.0) have been used for both the integrations. The state vectors at the reference epoch have been retrieved from the NASA JPL Horizons system. The integrations have been performed in the  ICRF/J2000.0 reference frame, with the ecliptic and mean equinox of the reference epoch, centered at the Solar System Barycenter (SSB). }\lb{EMB_Venus_astring}
\end{figure}
With its nominal maximum span of 3 m (peak-to-peak amplitude), also in this case such a perturbation would be detectable  with a cm-level ranging, and may pose some problems to the other signals of interest previously examined. Indeed, its mismodeled signature would impact the Lense-Thirring one at $4.5\times 10^{-1}$ level, while the bias on $J_2$ and the Schwarzschild effect is $2.3\times 10^{-2}$ and $7.5\times 10^{-6}$, respectively. Anyway, the time signatures are different.
Figure \ref{EMB_Venus_CePaVe} shows
the nominal perturbation on the Venus range by Ceres, Pallas, Vesta.
 The  peak-to-peak amplitude
 is 175 m, measurable at a $6\times 10^{-4}$ level with a ranging device accurate to 10-cm. The aliasing effect of the current mismodeling in $J_2$ and in the unmodeled Lense-Thirring effect is of the order of $2\times 10^{-2}, 1\times 10^{-2}$, respectively. Conversely, it turns out that the peak-to-peak amplitude of the mismodeled signature of Ceres, Pallas, Vesta is 1 m. It would largely alias the Lense-Thirring effect, while the systematic relative uncertainty induced on the gravitoelectric and $J_2$ range perturbations would be $8\times 10^{-6},2\times 10^{-2}$, respectively.
\begin{figure}[pb]
\centerline{\psfig{file=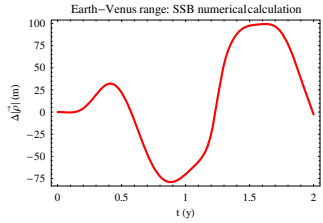
 ,width=4.7cm}}\vspace*{8pt}\caption{Difference $\Delta |\vec{\rho}|\doteq |\vec{\rho}_{\rm P}|-|\vec{\rho}_{\rm R}|$ in the numerically integrated EMB-Venus ranges with and without the nominal perturbation due to\protect\cite{CePaVe} Ceres, Pallas, Vesta  over $\Delta t=2$ yr. The same initial conditions (J2000.0) have been used for both the integrations. The state vectors at the reference epoch have been retrieved from the NASA JPL Horizons system. The integrations have been performed in the  ICRF/J2000.0 reference frame, with the ecliptic and mean equinox of the reference epoch, centered at the Solar System Barycenter (SSB). }\lb{EMB_Venus_CePaVe}
\end{figure}
\subsection{The Trans-Neptunian Objects}
Figure \ref{EMB_Venus_tnoring} shows the effect of the TNOs on the Venus range. Its peak-to-peak amplitude is about 50 cm: it may be detectable.
\begin{figure}[pb]
\centerline{\psfig{file=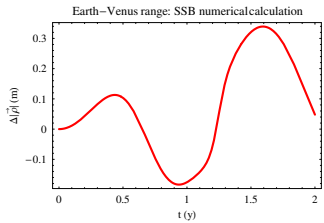
 ,width=4.7cm}}\vspace*{8pt}\caption{Difference $\Delta |\vec{\rho}|\doteq |\vec{\rho}_{\rm P}|-|\vec{\rho}_{\rm R}|$ in the numerically integrated EMB-Venus ranges with and without the nominal perturbation due to the ring of Trans-Neptunian Objects with\protect\cite{Pit} $m_{\rm ring}=5.26\times 10^{-8}$M$_{\odot}$ and $R_{\rm ring}=43$ a.u. over $\Delta t=2$ yr. The same initial conditions (J2000.0) have been used for both the integrations. The state vectors at the reference epoch have been retrieved from the NASA JPL Horizons system. The integrations have been performed in the  ICRF/J2000.0 reference frame, with the ecliptic and mean equinox of the reference epoch, centered at the Solar System Barycenter (SSB). }\lb{EMB_Venus_tnoring}
\end{figure}
Its bias on the Schwarzschild, $J_2$ and Lense-Thirring signals is $4\times 10^{-6}, 1.2\times 10^{-2}, 2.5\times 10^{-1}$, respectively. Concerning the aliasing effect on the gravitomagnetic effect, it must be noted that the temporal evolution of the two signals is different. This would help in decorrelating them.
\subsection{Violation of the Strong Equivalence Principle}
Figure \ref{EMB_Venus_eta} illustrates the nominal perturbation of the venusian range due to a SEP violation with $\eta=10^{-5}$.
\begin{figure}[pb]
\centerline{\psfig{file=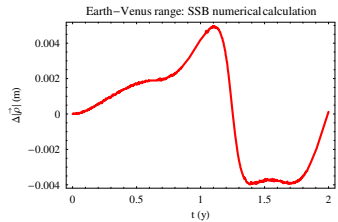
 ,width=4.7cm}}\vspace*{8pt}\caption{Difference $\Delta |\vec{\rho}|\doteq |\vec{\rho}_{\rm P}|-|\vec{\rho}_{\rm R}|$ in the numerically integrated EMB-Venus ranges with and without the nominal perturbation due to a violation of the Strong Equivalence Principle for $\eta=10^{-5}$  over $\Delta t=2$ yr. The same initial conditions (J2000.0) have been used for both the integrations. The state vectors at the reference epoch have been retrieved from the NASA JPL Horizons system. The integrations have been performed in the  ICRF/J2000.0 reference frame, with the ecliptic and mean equinox of the reference epoch, centered at the Solar System Barycenter (SSB). }\lb{EMB_Venus_eta}
\end{figure}
Such a signal is completely negligible because its peak-to-peak amplitude is of just 8 mm. Apart from the fact that it would be undetectable, it would be overwhelmed by all the other signatures, modeled or not, previously considered.
\subsection{Secular variation of the Newtonian constant of gravitation}
The effect of a secular variation of $G$ as large as $\dot G/G=-5.9\times 10^{-14}$ yr$^{-1}$ is depicted in Figure \ref{EMB_Venus_Gdot}.
\begin{figure}[pb]
\centerline{\psfig{file=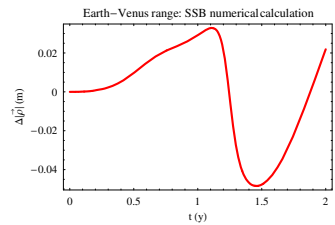
 ,width=4.7cm}}\vspace*{8pt}\caption{Difference $\Delta |\vec{\rho}|\doteq |\vec{\rho}_{\rm P}|-|\vec{\rho}_{\rm R}|$ in the numerically integrated EMB-Venus ranges with and without the nominal perturbation due to a secular variation of $G$ as large as\protect\cite{Pit} $\dot G/G=-5.9\times 10^{-14}$ yr$^{-1}$  over $\Delta t=2$ yr. The same initial conditions (J2000.0) have been used for both the integrations. The state vectors at the reference epoch have been retrieved from the NASA JPL Horizons system. The integrations have been performed in the  ICRF/J2000.0 reference frame, with the ecliptic and mean equinox of the reference epoch, centered at the Solar System Barycenter (SSB). }\lb{EMB_Venus_Gdot}
\end{figure}
Its peak-to-peak amplitude is about 7 cm, which is hardly detectable even with a cm-level ranging system. Moreover, such a signature would be easily biased by the other dynamical effects considered; for example, recall that the mismodeled effect of Ceres, Pallas, Vesta is as large as  1 m.
\subsection{The Pioneer Anomaly}
Figure \ref{EMB_Venus_Pio} illustrates the Venus range perturbation induced by the indirect effect of the Pioneer anomaly assumed acting only on the outer planets of the solar system. Their motions would be altered with respect to the standard case, thus reflecting also on the Earth-Venus distance.
\begin{figure}[pb]
\centerline{\psfig{file=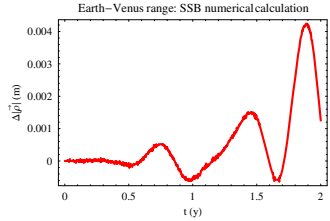
 ,width=4.7cm}}\vspace*{8pt}\caption{Difference $\Delta |\vec{\rho}|\doteq |\vec{\rho}_{\rm P}|-|\vec{\rho}_{\rm R}|$ in the numerically integrated EMB-Venus ranges with and without the nominal perturbation due to a Pioneer-like constant and unform radial acceleration of\protect\cite{Pio} $A_{\rm Pio}=8.74\times 10^{-10}$ m s$^{-2}$ acting upon Uranus, Neptune, Pluto, Eris over $\Delta t=2$ yr. The same initial conditions (J2000.0) have been used for both the integrations. The state vectors at the reference epoch have been retrieved from the NASA JPL Horizons system. The integrations have been performed in the  ICRF/J2000.0 reference frame, with the ecliptic and mean equinox of the reference epoch, centered at the Solar System Barycenter (SSB). }\lb{EMB_Venus_Pio}
\end{figure}
Its peak-to-peak amplitude amounts to 5 mm, about equal to the corresponding effect for Mercury (4 mm). It is negligible because it would be undetectable, given the expected cm-level accuracy of future interplanetary ranging devices.
 Moreover, the much larger aliasing effects of the other competing dynamical signatures would completely overwhelm it.
\subsection{Planet X}
More interesting is the situation for a putative planet X. Indeed, the peak-to-peak amplitude of its signal, illustrated in Figure \ref{EMB_Venus_X}-Figure \ref{EMB_Venus_X2}, is $3-5$ m.
\begin{figure}[pb]
\centerline{\psfig{file=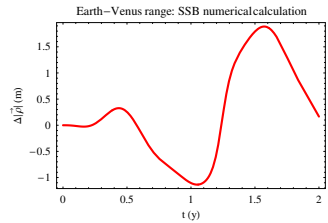
 ,width=4.7cm}}\vspace*{8pt}\caption{Difference $\Delta |\vec{\rho}|\doteq |\vec{\rho}_{\rm P}|-|\vec{\rho}_{\rm R}|$ in the numerically integrated EMB-Venus ranges with and without the perturbation due to hypothetical remote planet X lying almost in the ecliptic with minimum tidal parameter\protect\cite{Iorio} ${\mathcal{K}}_{\rm X}=1.6\times 10^{-26}$ s$^{-2}$  over $\Delta t=2$ yr. The same initial conditions (J2000.0) have been used for both the integrations. The state vectors at the reference epoch have been retrieved from the NASA JPL Horizons system. The integrations have been performed in the  ICRF/J2000.0 reference frame, with the ecliptic and mean equinox of the reference epoch, centered at the Solar System Barycenter (SSB). }\lb{EMB_Venus_X}
\end{figure}
\begin{figure}[pb]
\centerline{\psfig{file=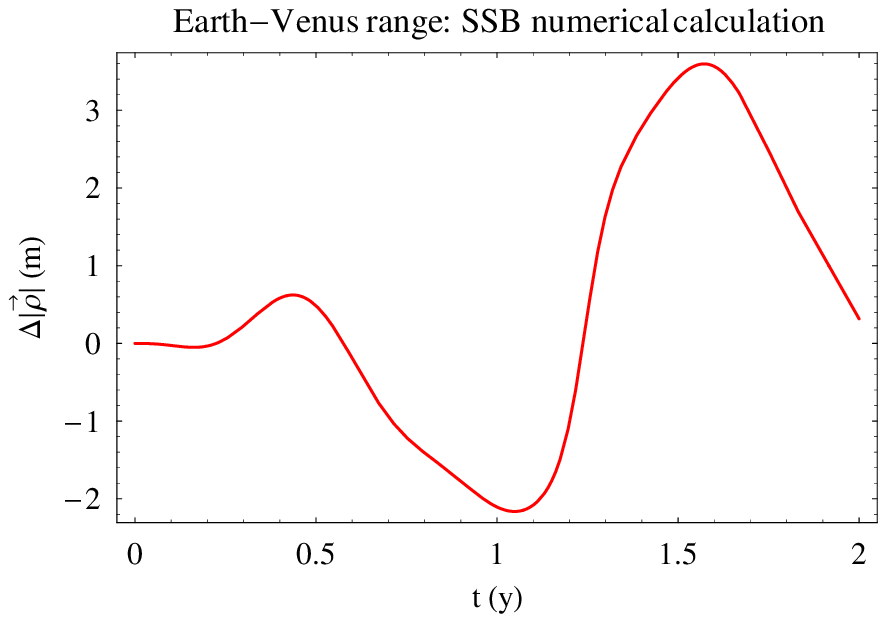
 ,width=4.7cm}}\vspace*{8pt}\caption{Difference $\Delta |\vec{\rho}|\doteq |\vec{\rho}_{\rm P}|-|\vec{\rho}_{\rm R}|$ in the numerically integrated EMB-Venus ranges with and without the perturbation due to hypothetical remote planet X lying almost in the ecliptic with maximum tidal parameter\protect\cite{Iorio} ${\mathcal{K}}_{\rm X}=2.7\times 10^{-26}$ s$^{-2}$  over $\Delta t=2$ yr. The same initial conditions (J2000.0) have been used for both the integrations. The state vectors at the reference epoch have been retrieved from the NASA JPL Horizons system. The integrations have been performed in the  ICRF/J2000.0 reference frame, with the ecliptic and mean equinox of the reference epoch, centered at the Solar System Barycenter (SSB). }\lb{EMB_Venus_X2}
\end{figure}
It would be detectable in future if a cm-level accuracy in ranging will be achieved. For the moment, the X's signature is compatible with the currently available range-residuals. Note that its pattern would be similar to that due to the TNOs ring, but its magnitude would be up to 10 times larger. On the contrary, the Lense-Thirring effect has a comparable, m-level size, but a different time signature.
\section{Earth-Mars range}\lb{mars}
For Mars we have at our disposal long time series of range residuals accurate to about $1-10$ m-level thanks to several spacecraft (Viking, Mars Pathfinder, Mars Global Surveyor, Mars Odyssey, Mars Reconnaissance Orbiter, Mars Express) which have orbited, or are still orbiting, the red planet. Figure B-10 of Ref.~\refcite{DE421} depicts the 1-way range residuals of the Viking Lander at Mars spanning from 1976 to 1982; they are at approximately 20 m level. Figure B-11 of  Ref.~\refcite{DE421} shows the 1-way range residuals of several post-Viking spacecraft; they generally cover a few years and are accurate to $5-10$ m.
\textcolor{black}{In the following, we will adopt an integration time span $\Delta t=5$ yr.}
\subsection{The Schwarzschild field of the Sun}
The general relativistic Schwarzschild perturbation of the Mars range is shown in Figure \ref{EMB_Mars_Schwa}.
\begin{figure}[pb]
\centerline{\psfig{file=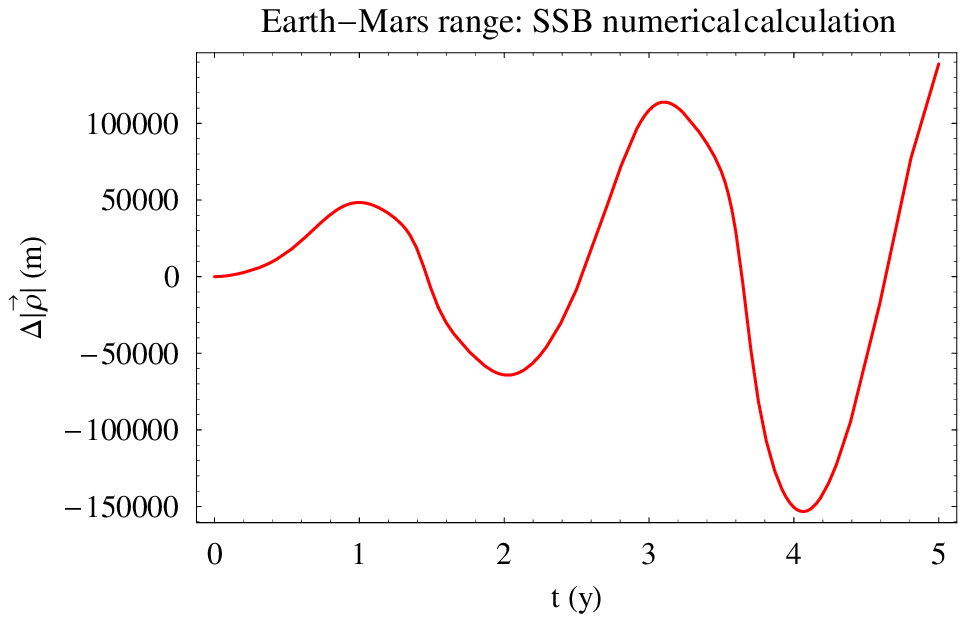
 ,width=4.7cm}}\vspace*{8pt}\caption{Difference $\Delta |\vec{\rho}|\doteq |\vec{\rho}_{\rm P}|-|\vec{\rho}_{\rm R}|$ in the numerically integrated EMB-Mars ranges with and without the perturbation due to the Sun's Schwarzschild  field over $\Delta t=\textcolor{black}{5}$ yr. The same initial conditions (J2000.0) have been used for both the integrations. The state vectors at the reference epoch have been retrieved from the NASA JPL Horizons system. The integrations have been performed in the  ICRF/J2000.0 reference frame, with the ecliptic and mean equinox of the reference epoch, centered at the Solar System Barycenter (SSB). }\lb{EMB_Mars_Schwa}
\end{figure}
Its peak-to-peak amplitude is \textcolor{black}{$2.5\times 10^5$} m. Thus, a $5-10$ cm-level ranging device operating continuously over \textcolor{black}{5} years would allow a relative accuracy in measuring it of  \textcolor{black}{$2-4\times 10^{-7}$}. Also in this case, several potential sources of systematic errors are to be carefully considered.
\subsection{The oblateness of the Sun}
Figure \ref{EMB_Mars_J2} illustrates the nominal signal due to the Sun's quadrupole mass moment for $J_2=2\times 10^{-7}$ computed in a frame aligned with the Sun's equator.
\begin{figure}[pb]
\centerline{\psfig{file=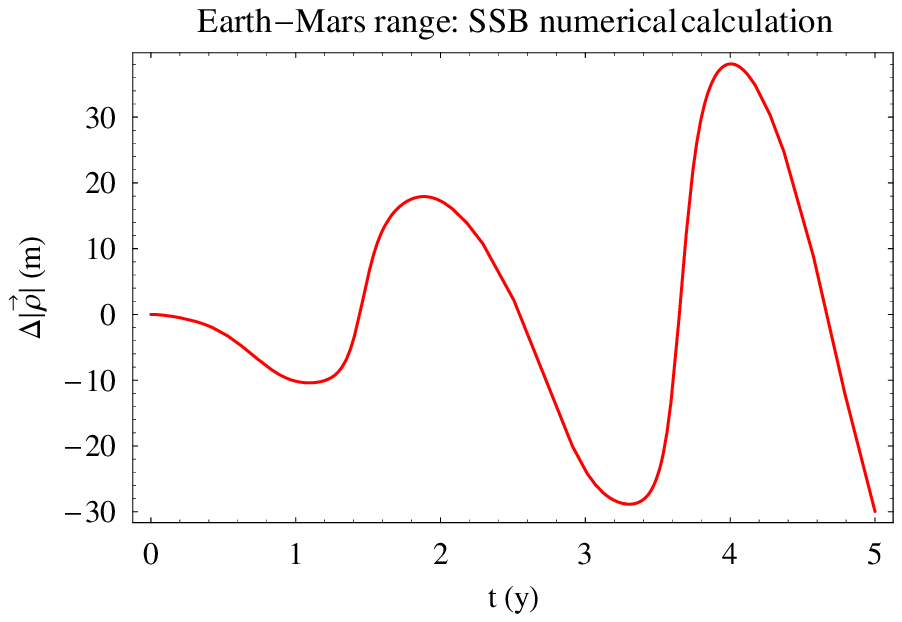
 ,width=4.7cm}}\vspace*{8pt}\caption{Difference $\Delta |\vec{\rho}|\doteq |\vec{\rho}_{\rm P}|-|\vec{\rho}_{\rm R}|$ in the numerically integrated EMB-Mars ranges with and without the nominal perturbation due to the Sun's quadrupole mass moment $J_2=2.0\times 10^{-7}$ over $\Delta t=\textcolor{black}{5}$ yr. The same initial conditions (J2000.0) have been used for both the integrations. The state vectors at the reference epoch have been retrieved from the NASA JPL Horizons system. The integrations have been performed in the  ICRF/J2000.0 reference frame, with the mean equinox of the reference epoch and the reference $\{xy\}$ plane rotated from the mean ecliptic of the epoch to the Sun's equator, centered at the Solar System Barycenter (SSB). }\lb{EMB_Mars_J2}
\end{figure}
Its peak-to-peak amplitude amounts to about \textcolor{black}{70} m; thus, its effect would be well measurable at a \textcolor{black}{$7\times 10^{-4}-1\times 10^{-3}$} level by means of a new ranging facility with an accuracy of the order of cm. Concerning its actual presence in the present-day range residuals, it must be noted that the dynamical action of the solar $J_2$ has always been modeled in producing them. Since $J_2$ is nowadays accurate to $10^{-1}$, the corresponding mismodeled signature would be as large as about \textcolor{black}{$7$} m, i.e. well compatible with the range residuals available.
Its impact as a source of systematic uncertainty on the Schwarzschild signal amounts to \textcolor{black}{$3\times 10^{-5}$}; note, however, the different time signatures of Figure \ref{EMB_Mars_Schwa} and Figure  \ref{EMB_Mars_J2}.
\subsection{The Lense-Thirring effect of the Sun}
The Lense-Thirring range perturbation, computed in a frame aligned with the Sun's equator, is shown in Figure \ref{EMB_Mars_LT}.
\begin{figure}[pb]
\centerline{\psfig{file=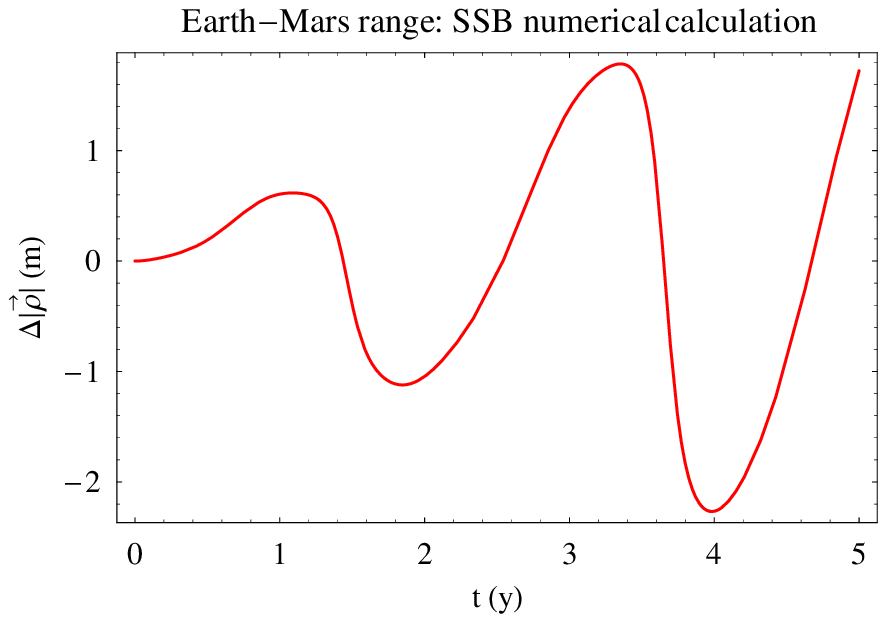
 ,width=4.7cm}}\vspace*{8pt}\caption{Difference $\Delta |\vec{\rho}|\doteq |\vec{\rho}_{\rm P}|-|\vec{\rho}_{\rm R}|$ in the numerically integrated EMB-Mars ranges with and without the perturbation due to the Sun's Lense-Thirring field over $\Delta t=\textcolor{black}{5}$ yr. The same initial conditions (J2000.0) have been used for both the integrations. The state vectors at the reference epoch have been retrieved from the NASA JPL Horizons system. The integrations have been performed in the  ICRF/J2000.0 reference frame, with the mean equinox of the reference epoch and the reference $\{xy\}$ plane rotated from the mean ecliptic of the epoch to the Sun's equator, centered at the Solar System Barycenter (SSB). }\lb{EMB_Mars_LT}
\end{figure}
Its peak-to-peak amplitude is about \textcolor{black}{4} m, not too far from the present-day range accuracy; thus, its existence as predicted by general relativity is not in contrast with the range residuals currently available. It could be measured with a future cm-level ranging system at a \textcolor{black}{$1-2.5\%$} level.
If not properly modeled, the gravitomagnetic signature would impact the Schwarzschild one at a \textcolor{black}{$1.6\times 10^{-5}$} level; moreover, note the similar time evolution of the two signals. Concerning $J_2$, the Lense-Thirring effect would bias its signal at a $6\%$ level. Conversely, if one looks at $J_2$ as a potential source of systematic bias for the recovery of the gravitomagnetic effect, the mismodeled signature of the Sun's quadrupolar mass moment would be \textcolor{black}{$1.7$} times larger than it. An improvement in its knowledge by one order of magnitude, as expected from, e.g., BepiColombo, would push its bias on the Lense-Thirring signal at \textcolor{black}{$17\%$}. Anyway, it must be noted that their temporal evolutions are different.
\subsection{The ring of the minor asteroids and Ceres, Pallas and Vesta}
The range perturbation of the ring of the minor asteroids is reproduced in Figure \ref{EMB_Mars_tnoring}.
\begin{figure}[pb]
\centerline{\psfig{file=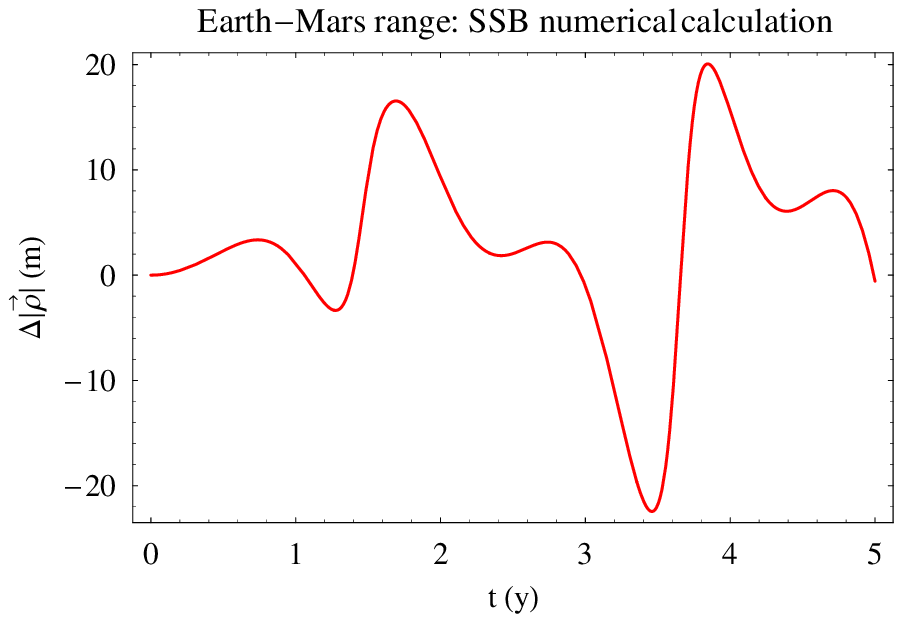
 ,width=4.7cm}}\vspace*{8pt}\caption{Difference $\Delta |\vec{\rho}|\doteq |\vec{\rho}_{\rm P}|-|\vec{\rho}_{\rm R}|$ in the numerically integrated EMB-Mars ranges with and without the nominal perturbation due to the ring of minor asteroids with\protect\cite{Fienga} $m_{\rm ring}=1\times 10^{-10}$M$_{\odot}$  and $R_{\rm ring}=3.14$ a.u. over $\Delta t=\textcolor{black}{5}$ yr. The same initial conditions (J2000.0) have been used for both the integrations. The state vectors at the reference epoch have been retrieved from the NASA JPL Horizons system. The integrations have been performed in the  ICRF/J2000.0 reference frame, with the ecliptic and mean equinox of the reference epoch, centered at the Solar System Barycenter (SSB). }\lb{EMB_Mars_astring}
\end{figure}
Its nominal peak-to-peak amplitude is \textcolor{black}{40} m; by considering a $30\%$ uncertainty in the mass of such a ring, the related mismodeled signal would be as large as \textcolor{black}{12} m. It would impact the recovery of the Schwarzschild and $J_2$  signals at \textcolor{black}{$5\times 10^{-5}, 1.7\times 10^{-1}$} level, respectively, while the Lense-Thirring signature would be swamped.
Note that the time signature of the minor asteroids is different from the relativistic ones and more similar to that due to $J_2$.

 The nominal perturbation on the  range of Mars by Ceres, Pallas, Vesta is
shown  in Figure \ref{EMB_Mars_CePaVe}.
\begin{figure}[pb]
\centerline{\psfig{file=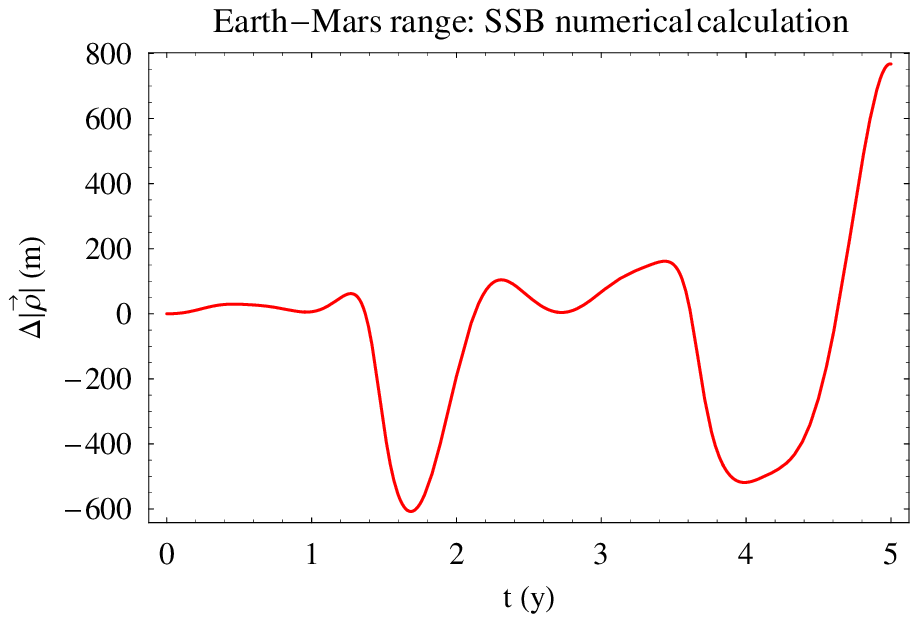
 ,width=4.7cm}}\vspace*{8pt}\caption{Difference $\Delta |\vec{\rho}|\doteq |\vec{\rho}_{\rm P}|-|\vec{\rho}_{\rm R}|$ in the numerically integrated EMB-Mars ranges with and without the nominal perturbation due to\protect\cite{CePaVe} Ceres, Pallas, Vesta  over $\Delta t=\textcolor{black}{5}$ yr. The same initial conditions (J2000.0) have been used for both the integrations. The state vectors at the reference epoch have been retrieved from the NASA JPL Horizons system. The integrations have been performed in the  ICRF/J2000.0 reference frame, with the ecliptic and mean equinox of the reference epoch, centered at the Solar System Barycenter (SSB). }\lb{EMB_Mars_CePaVe}
\end{figure}
\textcolor{black}{ For previous analytical and numerical investigations of the their impact on the motions of the Earth and Mars, see Ref.~\refcite{scazz}. }
 The  peak-to-peak amplitude is as large as \textcolor{black}{1400} m, measurable at a \textcolor{black}{$7\times 10^{-5}$} level by assuming an accuracy of 10 cm in the ranging device. The aliasing effect of the present-day mismodeling in $J_2$ and in the unmodeled Lense-Thirring effect is of the order of \textcolor{black}{$5\times 10^{-3}, 2.8\times 10^{-3}$}, respectively. On the other hand, it can be shown that the peak-to-peak amplitude of the mismodeled signature of the main asteroids considered here is up to \textcolor{black}{14} m. It would overwhelm the Lense-Thirring effect, while the systematic relative uncertainty on the Schwarzschild and $J_2$ range perturbations would be \textcolor{black}{$5.6\times 10^{-5},2\times 10^{-1}$}, respectively.

\textcolor{black}{ At the end, it must be remarked that the present analysis of the impact of asteroids on Mars should be regarded just as necessarily incomplete and preliminary. Indeed, over timescales of the order of its orbital period\footnote{ It amounts to about $1.9$ yr, i.e. it is comparable to the expected time span for PLR operations. }, the orbit of Mars is notably affected by a larger number of different minor bodies\cite{aste1,aste2,aste3}. On the contrary, appreciable effects of them on Mercury and Venus occur on timescales some decades long\cite{lungo1,lungo2}. Modern ephemerides like\cite{Fienga08,Fienga09}, e.g, INPOP06-INPOP08 include the gravitational influences of up to 300 most perturbing asteroids of the Martian orbit. Limiting to a ring model may lead to inaccuracies. Anyway, recent developments\cite{kucy} may have somewhat mitigated such a risk, especially over timescales of a few years.
Accurate determinations of the masses of about hundred minor asteroids\cite{gaia1,gaia2,gaia3,gaia4} is one of the main goals of  GAIA\footnote{It is an astrometric spacecraft-based ESA mission whose launch is scheduled for 2012. See on the WEB http://sci.esa.int/science-e/www/area/index.cfm?fareaid=26.} \cite{gaia}.
Isolating the gravitational perturbation caused by a single asteroid on Mars, being strongly correlated and mixed up with those of many other asteroids, is a non trivial task. It is outside the scopes of the present paper, being a possible subject for further, dedicated analyses by independent teams of skilful astronomers. }
\subsection{The Trans-Neptunian Objects}
More serious is the effect on the Mars range of the ring of the TNOs: it is shown in Figure \ref{EMB_Mars_tnoring}.
\begin{figure}[pb]
\centerline{\psfig{file=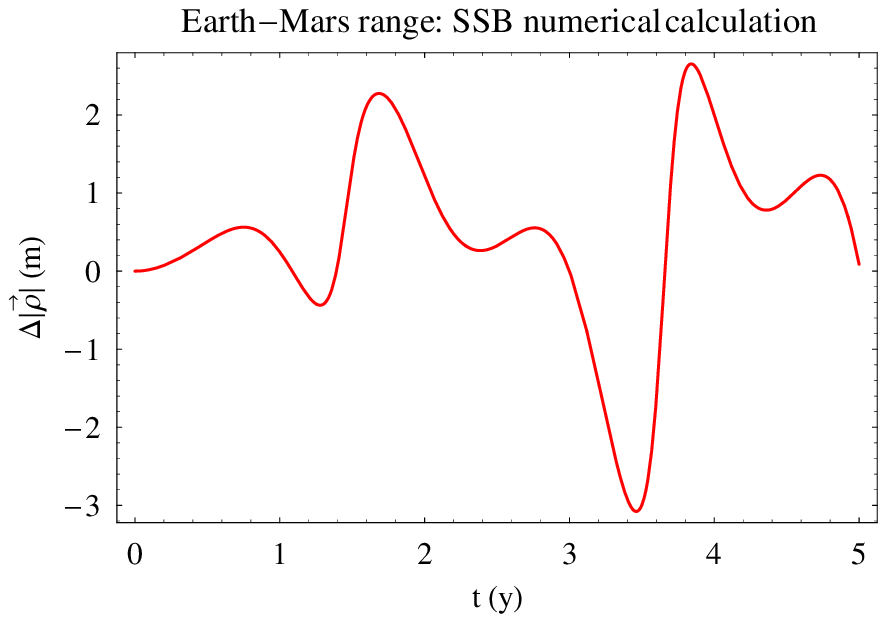
 ,width=4.7cm}}\vspace*{8pt}\caption{Difference $\Delta |\vec{\rho}|\doteq |\vec{\rho}_{\rm P}|-|\vec{\rho}_{\rm R}|$ in the numerically integrated EMB-Mars ranges with and without the nominal perturbation due to the ring of Trans-Neptunian Objects with\protect\cite{Pit}  $m_{\rm ring}=5.26\times 10^{-8}$M$_{\odot}$  and $R_{\rm ring}=43$ a.u. over $\Delta t=\textcolor{black}{5}$ yr. The same initial conditions (J2000.0) have been used for both the integrations. The state vectors at the reference epoch have been retrieved from the NASA JPL Horizons system. The integrations have been performed in the  ICRF/J2000.0 reference frame, with the ecliptic and mean equinox of the reference epoch, centered at the Solar System Barycenter (SSB). }\lb{EMB_Mars_tnoring}
\end{figure}
Its peak-to-peak amplitude is about \textcolor{black}{$5$} m, compatible with the present-day range residuals available. Thus, the TNOs, whose mass should  be conservatively considered as uncertain at a $100\%$ level,  would impact the Schwarzschild signal at \textcolor{black}{$2\times 10^{-5}$} level: the signatures are different. The $J_2$ effect would be biased at a \textcolor{black}{$7\%$} level, while the Lense-Thirring one would be overwhelmed by the TNOs, although their patterns are not equal.
\subsection{Violation of the Strong Equivalence Principle}
Moving to exotic effects, a violation if SEP driven by $\eta=10^{-5}$ is shown in Figure \ref{EMB_Mars_eta}.
 \begin{figure}[pb]
\centerline{\psfig{file=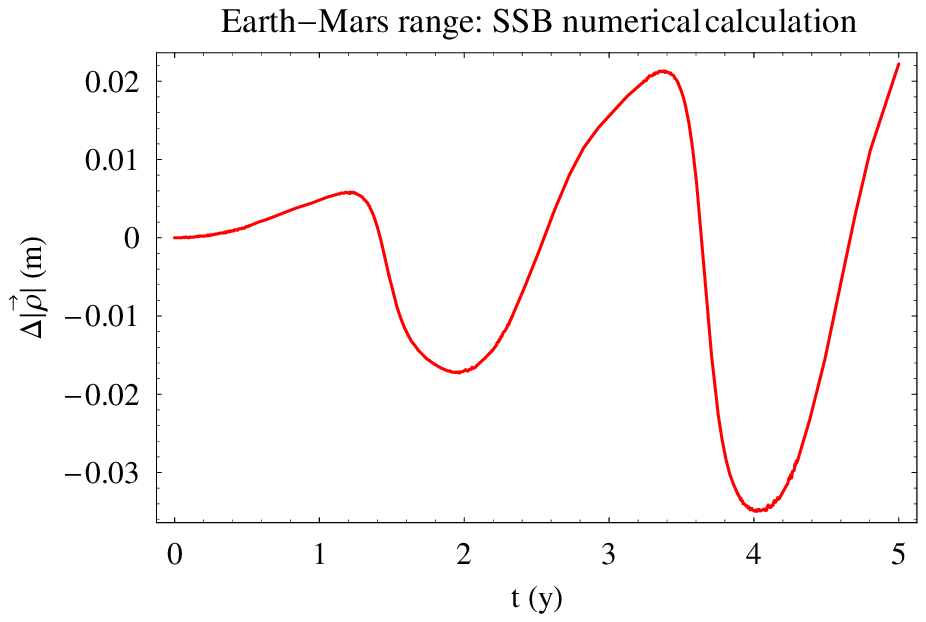
 ,width=4.7cm}}\vspace*{8pt}\caption{Difference $\Delta |\vec{\rho}|\doteq |\vec{\rho}_{\rm P}|-|\vec{\rho}_{\rm R}|$ in the numerically integrated EMB-Mars ranges with and without the nominal perturbation due a violation of SEP for $\eta=10^{-5}$ over $\Delta t=\textcolor{black}{5}$ yr. The same initial conditions (J2000.0) have been used for both the integrations. The state vectors at the reference epoch have been retrieved from the NASA JPL Horizons system. The integrations have been performed in the  ICRF/J2000.0 reference frame, with the ecliptic and mean equinox of the reference epoch, centered at the Solar System Barycenter (SSB). }\lb{EMB_Mars_eta}
\end{figure}
Its peak-to-peak amplitude is about \textcolor{black}{5} cm. Also in this case, it would be undetectable, and it would be totally swamped by the other dynamical effects considered, modeled or not.
\subsection{Secular variation of the Newtonian constant of gravitation}
The effect on the Mars range of a variation of $G$ as large as that determined by Pitjeva in Ref.~\refcite{Pit} is reproduced in Figure \ref{EMB_Mars_Gdot}.
\begin{figure}[pb]
\centerline{\psfig{file=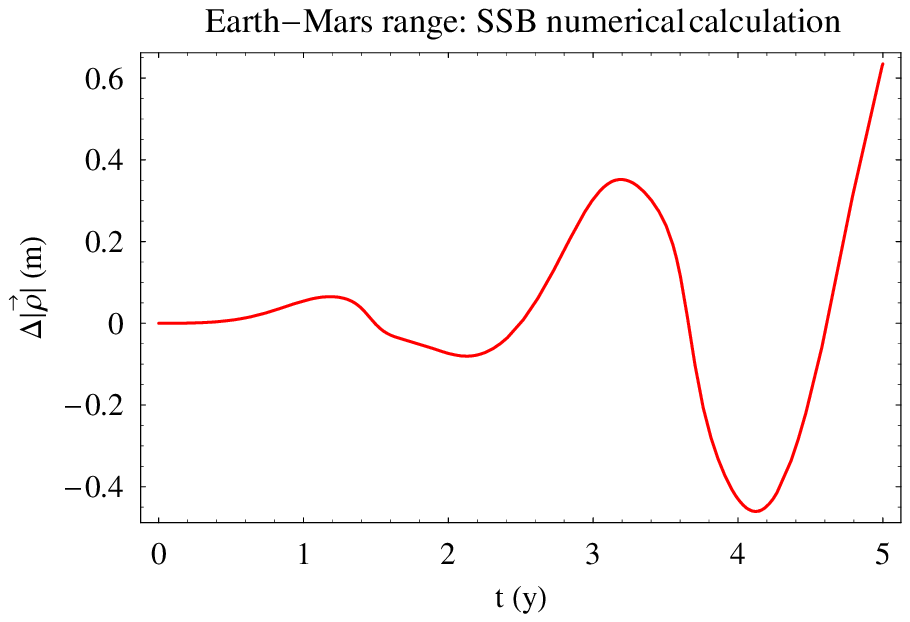
 ,width=4.7cm}}\vspace*{8pt}\caption{Difference $\Delta |\vec{\rho}|\doteq |\vec{\rho}_{\rm P}|-|\vec{\rho}_{\rm R}|$ in the numerically integrated EMB-Mars ranges with and without the nominal perturbation due to a secular variation of $G$ as large as\protect\cite{Pit} $\dot G/G=-5.9\times 10^{-14}$ yr$^{-1}$  over $\Delta t=\textcolor{black}{5}$ yr. The same initial conditions (J2000.0) have been used for both the integrations. The state vectors at the reference epoch have been retrieved from the NASA JPL Horizons system. The integrations have been performed in the  ICRF/J2000.0 reference frame, with the ecliptic and mean equinox of the reference epoch, centered at the Solar System Barycenter (SSB). }\lb{EMB_Mars_Gdot}
\end{figure}
It is as large as \textcolor{black}{1} m, i.e. it would be barely detectable because it would  likely be removed from the signal when fitting the initial conditions.
Moreover, it would be overwhelmed by the aliasing effects of all the other competing dynamical forces considered so far. Suffices it to say that the TNOs signal is \textcolor{black}{5} times larger; the present-day uncertainties in the masses of Ceres, Pallas and Vesta yield a signal of \textcolor{black}{up to 14} m.
\subsection{The Pioneer Anomaly}
Figure \ref{EMB_Mars_Pio} depicts the indirect effect of the Pioneer anomaly on the Mars range.
\begin{figure}[pb]
\centerline{\psfig{file=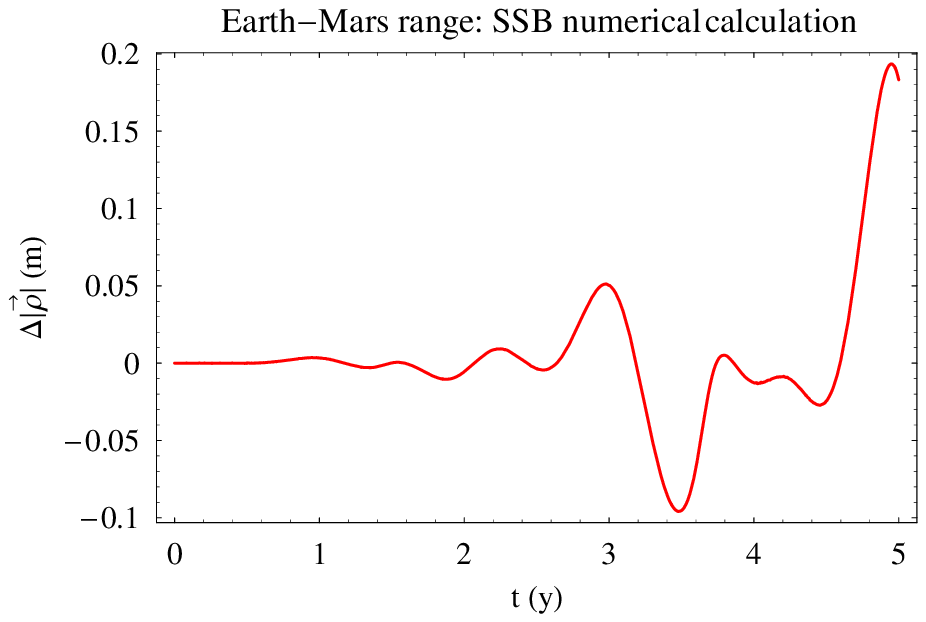
 ,width=4.7cm}}\vspace*{8pt}\caption{Difference $\Delta |\vec{\rho}|\doteq |\vec{\rho}_{\rm P}|-|\vec{\rho}_{\rm R}|$ in the numerically integrated EMB-Mars ranges with and without the nominal perturbation due to a Pioneer-like constant and unform radial acceleration of\protect\cite{Pio} $A_{\rm Pio}=8.74\times 10^{-10}$ m s$^{-2}$ acting upon Uranus, Neptune, Pluto, Eris over $\Delta t=\textcolor{black}{5}$ yr. The same initial conditions (J2000.0) have been used for both the integrations. The state vectors at the reference epoch have been retrieved from the NASA JPL Horizons system. The integrations have been performed in the  ICRF/J2000.0 reference frame, with the ecliptic and mean equinox of the reference epoch, centered at the Solar System Barycenter (SSB). }\lb{EMB_Mars_Pio}
\end{figure}
Its peak-to-peak amplitude is \textcolor{black}{30} cm.
Also in this case, such a potential exotic effect would likely be too small to be realistically detected even with a future,  advanced, cm-level ranging system.
The systematic bias due to the other standard Newtonian and relativistic effects would be largely overwhelming.
\subsection{Planet X}
Finally, we consider the action of planet X in Figure \ref{EMB_Mars_X}-Figure \ref{EMB_Mars_X2}.
\begin{figure}[pb]
\centerline{\psfig{file=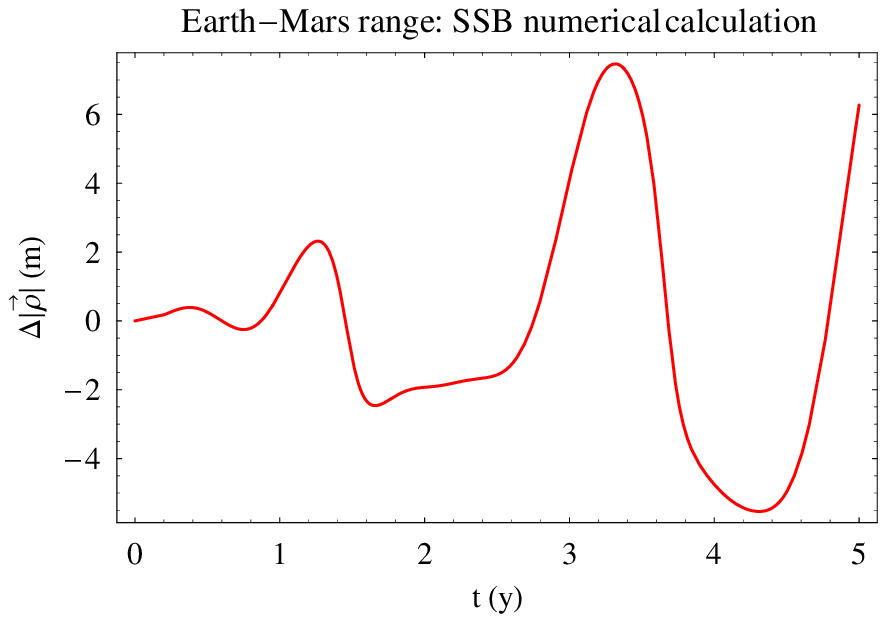
 ,width=4.7cm}}\vspace*{8pt}\caption{Difference $\Delta |\vec{\rho}|\doteq |\vec{\rho}_{\rm P}|-|\vec{\rho}_{\rm R}|$ in the numerically integrated EMB-Mars ranges with and without the perturbation due to hypothetical remote planet X lying almost in the ecliptic with minimum tidal parameter\protect\cite{Iorio} ${\mathcal{K}}_{\rm X}=1.6\times 10^{-26}$ s$^{-2}$  over $\Delta t=\textcolor{black}{5}$ yr. The same initial conditions (J2000.0) have been used for both the integrations. The state vectors at the reference epoch have been retrieved from the NASA JPL Horizons system. The integrations have been performed in the  ICRF/J2000.0 reference frame, with the ecliptic and mean equinox of the reference epoch, centered at the Solar System Barycenter (SSB). }\lb{EMB_Mars_X}
\end{figure}
\begin{figure}[pb]
\centerline{\psfig{file=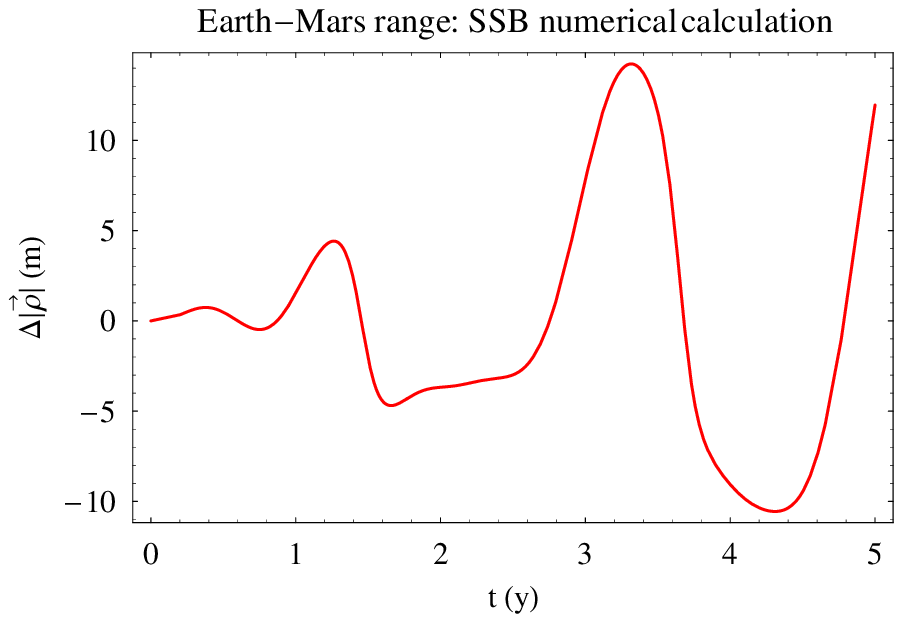
 ,width=4.7cm}}\vspace*{8pt}\caption{Difference $\Delta |\vec{\rho}|\doteq |\vec{\rho}_{\rm P}|-|\vec{\rho}_{\rm R}|$ in the numerically integrated EMB-Mars ranges with and without the perturbation due to hypothetical remote planet X lying almost in the ecliptic with maximum tidal parameter\protect\cite{Iorio} ${\mathcal{K}}_{\rm X}=2.7\times 10^{-26}$ s$^{-2}$  over $\Delta t=\textcolor{black}{5}$ yr. The same initial conditions (J2000.0) have been used for both the integrations. The state vectors at the reference epoch have been retrieved from the NASA JPL Horizons system. The integrations have been performed in the  ICRF/J2000.0 reference frame, with the ecliptic and mean equinox of the reference epoch, centered at the Solar System Barycenter (SSB). }\lb{EMB_Mars_X2}
\end{figure}
Its peak-to-peak amplitude is \textcolor{black}{$10-20$} m; it is not in contrast with the present-day range residuals from the Martian spacecraft, also because  X has not been explicitly modeled in producing them, and part of  its signature could have been removed from the signal in fitting the initial conditions.
Such a range perturbation could be measured with a future cm-level ranging system. The bias due to the TNOs would be \textcolor{black}{$2-4$} times smaller, and their signature would be different from that of X. Concerning the Lense-Thirring effect, if not modeled it would not mimic the action of X, and its magnitude would be \textcolor{black}{$2.5-5$} times smaller that that of X. The mismodeled signal due to the Sun's $J_2$ would be about \textcolor{black}{$1.4-3$} times smaller, but the $S/N$ ratio would become more favorable after the expected improvements by one order of magnitude in our knowledge of $J_2$.

\textcolor{black}{
\section{Earth-Jupiter range}\lb{jupiter}
Concerning  the Earth-Jupiter range, Figure B-14, pag. 18 of Ref.~\refcite{DE421} tells us that there are just a few sparse points from the encounters with the Pioneer 10/11, Voyager 1/2, Ulysses and Cassini probes whose accuracy is of a few km, apart from the Ulysses' point.\\
Jupiter will be orbited in the next years by the approved Juno spacecraft\cite{juno}: the nominal mission duration is 1 yr. Europa Jupiter System Mission, or Laplace, (Laplace/EJSM) is an  unmanned  mission\footnote{\textcolor{black}{See on the WEB http://sci.esa.int/science-e/www/object/index.cfm?fobjectid=44037.}} jointly proposed by NASA and ESA in the framework of the Cosmic Vision $2015-2025$ programme for the in-depth exploration of jovian  moons with a focus on Europa, Ganymede and the Jupiter's magnetosphere; a possible launch date would be in 2020. Continuous ranging to both the spacecraft would certainly improve our knowledge of the Earth-Jupiter distance, although it is not possible to give a precise figure for that. \\
 Determining accurately the motion of Jupiter may also have consequences on our precise knowledge of certain important features of the Cosmic Microwave Background (CMB) having implications on our views on dark matter and dark energy. Indeed, the pointing of the detectors  of the Wilkinson Microwave Anisotropy Probe (WMAP) is determined empirically from observations of Jupiter. Some results obtained in such a way have been recently questioned; see Refs.~\cite{wmap1,wmap2} and references therein about the ongoing debate.
\subsection{The Schwarzschild field of the Sun}
Figure \ref{EMB_Jupiter_Schwa} depicts the  perturbation caused by the solar Schwarzschild field on the Earth-Jupiter range.
\begin{figure}[pb]
\centerline{\psfig{file=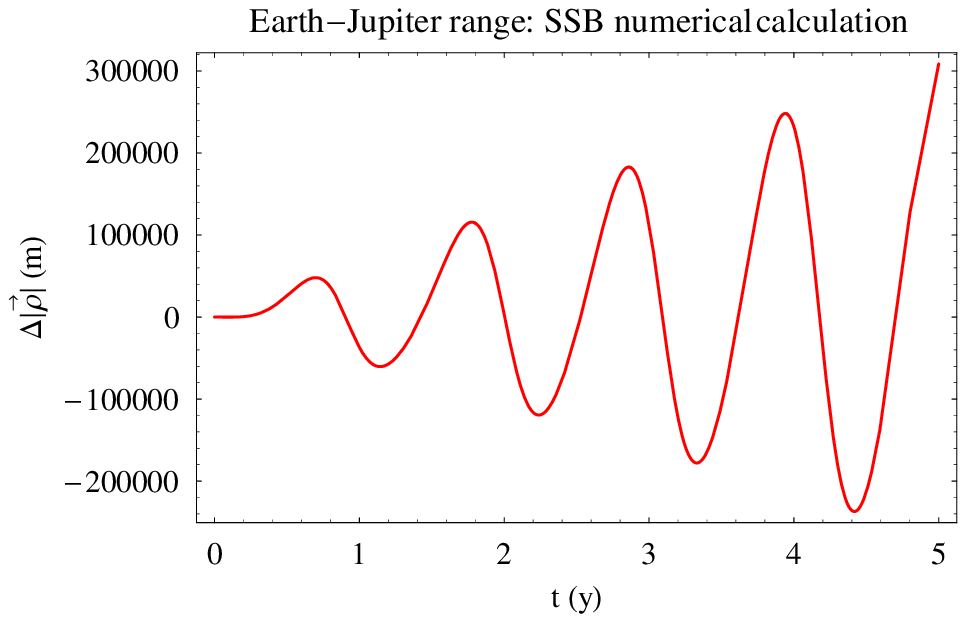
 ,width=4.7cm}}\vspace*{8pt}\caption{\textcolor{black}{Difference $\Delta |\vec{\rho}|\doteq |\vec{\rho}_{\rm P}|-|\vec{\rho}_{\rm R}|$ in the numerically integrated EMB-Jupiter ranges with and without the perturbation due to the Sun's Schwarzschild  field over $\Delta t=\textcolor{black}{5}$ yr. The same initial conditions (J2000.0) have been used for both the integrations. The state vectors at the reference epoch have been retrieved from the NASA JPL Horizons system. The integrations have been performed in the  ICRF/J2000.0 reference frame, with the ecliptic and mean equinox of the reference epoch, centered at the Solar System Barycenter (SSB).} }\lb{EMB_Jupiter_Schwa}
\end{figure}
The peak-to-peak nominal amplitude is  $\textcolor{black}{5}\times 10^5$ m.
\subsection{The oblateness of the Sun}
The effect of the solar oblateness on the jovian range is shown in Figure \ref{EMB_Jupiter_J2}.
\begin{figure}[pb]
\centerline{\psfig{file=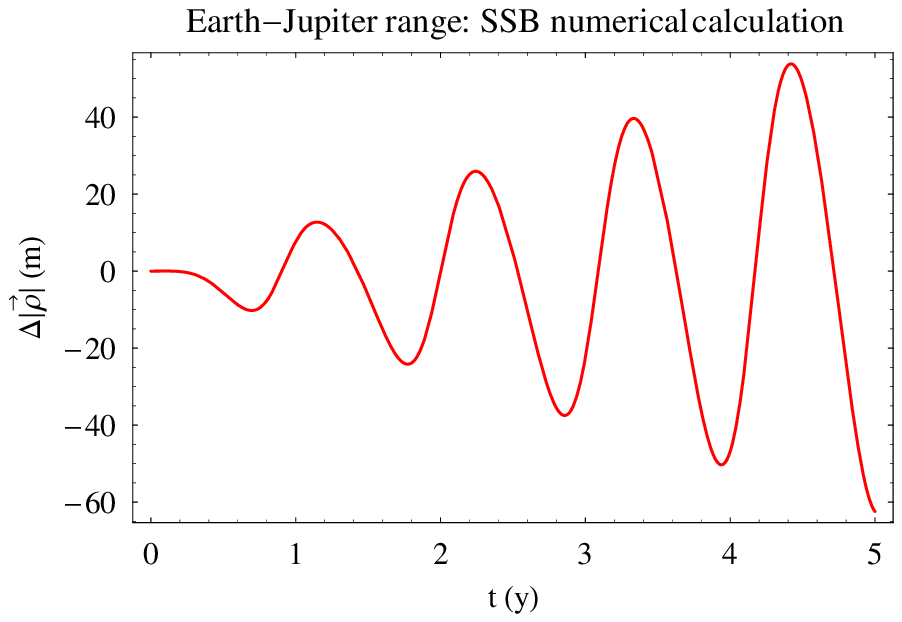
 ,width=4.7cm}}\vspace*{8pt}\caption{\textcolor{black}{Difference $\Delta |\vec{\rho}|\doteq |\vec{\rho}_{\rm P}|-|\vec{\rho}_{\rm R}|$ in the numerically integrated EMB-Jupiter ranges with and without the nominal perturbation due to the Sun's quadrupole mass moment $J_2=2.0\times 10^{-7}$ over $\Delta t=\textcolor{black}{5}$ yr. The same initial conditions (J2000.0) have been used for both the integrations. The state vectors at the reference epoch have been retrieved from the NASA JPL Horizons system. The integrations have been performed in the  ICRF/J2000.0 reference frame, with the mean equinox of the reference epoch and the reference $\{xy\}$ plane rotated from the mean ecliptic of the epoch to the Sun's equator, centered at the Solar System Barycenter (SSB). }}\lb{EMB_Jupiter_J2}
\end{figure}
The peak-to-peak nominal amplitude is of the order of more than \textcolor{black}{100} m. Given the current level of uncertainty in the solar $J_2$, its mismodeled effect amounts to $2\times 10^{-\textcolor{black}{4}}$ of the general relativistic Schwarzschild signal. Note that the temporal pattern is quite different with respect to the Schwarzschild one.
\subsection{The Lense-Thirring effect of the Sun}
Figure \ref{EMB_Jupiter_LT} shows the effect of the solar gravitomagnetic field on the Earth-Jupiter range.
\begin{figure}[pb]
\centerline{\psfig{file=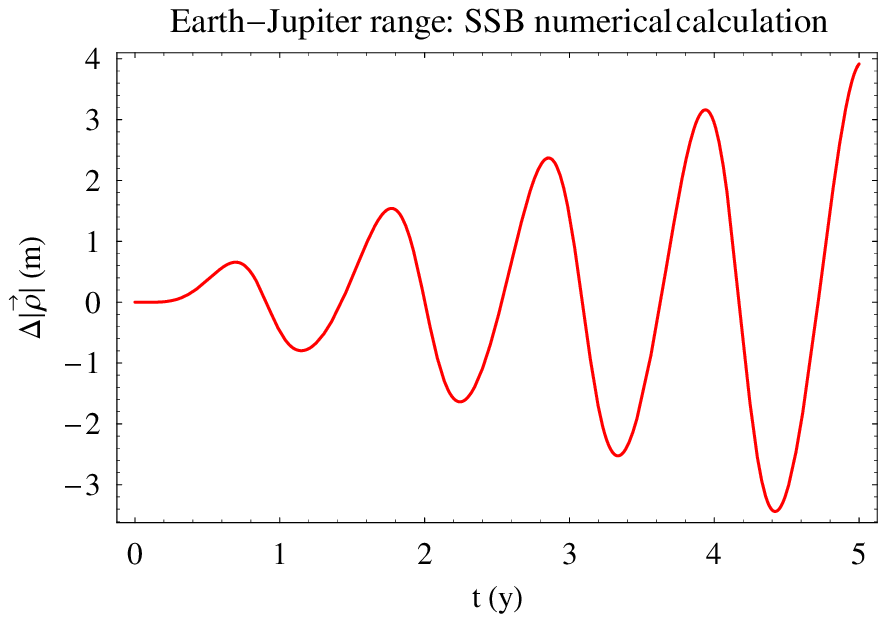
 ,width=4.7cm}}\vspace*{8pt}\caption{\textcolor{black}{Difference $\Delta |\vec{\rho}|\doteq |\vec{\rho}_{\rm P}|-|\vec{\rho}_{\rm R}|$ in the numerically integrated EMB-Jupiter ranges with and without the perturbation due to the Sun's Lense-Thirring field over $\Delta t=\textcolor{black}{5}$ yr. The same initial conditions (J2000.0) have been used for both the integrations. The state vectors at the reference epoch have been retrieved from the NASA JPL Horizons system. The integrations have been performed in the  ICRF/J2000.0 reference frame, with the mean equinox of the reference epoch and the reference $\{xy\}$ plane rotated from the mean ecliptic of the epoch to the Sun's equator, centered at the Solar System Barycenter (SSB).} }\lb{EMB_Jupiter_LT}
\end{figure}
Its peak-to-peak nominal amplitude is as large as about \textcolor{black}{7} m. Its time signature is different from that of $J_2^{\odot}$, whose mismodelled effect would be of the same order of magnitude, but is similar to that due to the Schwarzschild field. Conversely, the Lense-Thirring effect, if not accounted for, would bias the $J_2^{\odot}$ and Schwarzschild signals at $7\%$ and $10^{-5}$ level, respectively.
\subsection{The ring of the minor asteroids and Ceres, Pallas and Vesta}
The nominal range perturbation caused by the ring of minor asteroids is in Figure \ref{EMB_Jupiter_astring}.
\begin{figure}[pb]
\centerline{\psfig{file=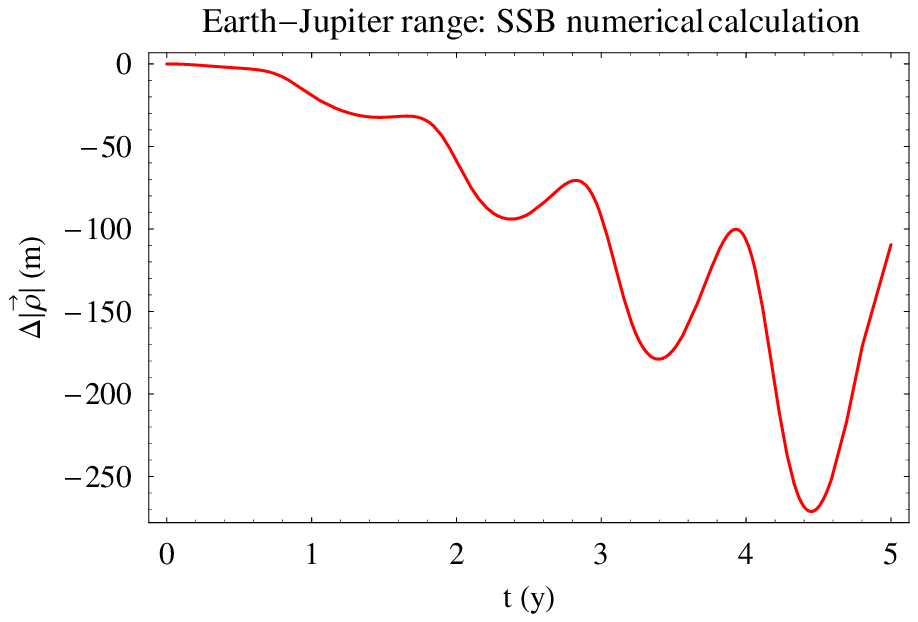
 ,width=4.7cm}}\vspace*{8pt}\caption{\textcolor{black}{Difference $\Delta |\vec{\rho}|\doteq |\vec{\rho}_{\rm P}|-|\vec{\rho}_{\rm R}|$ in the numerically integrated EMB-Jupiter ranges with and without the nominal perturbation due to the ring of minor asteroids with\protect\cite{Fienga} $m_{\rm ring}=1\times 10^{-10}$M$_{\odot}$  and $R_{\rm ring}=3.14$ a.u. over $\Delta t=\textcolor{black}{5}$ yr. The same initial conditions (J2000.0) have been used for both the integrations. The state vectors at the reference epoch have been retrieved from the NASA JPL Horizons system. The integrations have been performed in the  ICRF/J2000.0 reference frame, with the ecliptic and mean equinox of the reference epoch, centered at the Solar System Barycenter (SSB).} }\lb{EMB_Jupiter_astring}
\end{figure}
The peak-to-peak amplitude is \textcolor{black}{250} m; the mismodeled component would be as large as \textcolor{black}{75} m. It would overwhelm the Lense-Thirring signature; the alias on the Schwarzschild and $J_2$ signals would be $1\textcolor{black}{.5}\times 10^{-4}$ and $\textcolor{black}{75}\%$, respectively. Note that the temporal pattern of the asteroidal ring is quite different from that of the previous effects considered.\\
The effect of Ceres, Pallas and Vesta is illustrated in Figure \ref{EMB_Jupiter_CePaVe}.
\begin{figure}[pb]
\centerline{\psfig{file=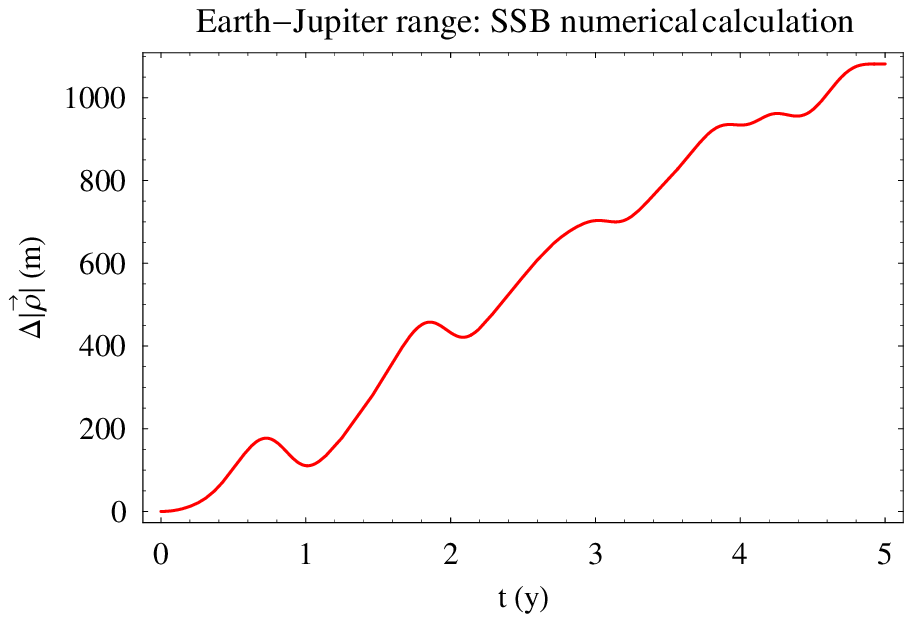
 ,width=4.7cm}}\vspace*{8pt}\caption{\textcolor{black}{Difference $\Delta |\vec{\rho}|\doteq |\vec{\rho}_{\rm P}|-|\vec{\rho}_{\rm R}|$ in the numerically integrated EMB-Jupiter ranges with and without the nominal perturbation due to\protect\cite{CePaVe}  Ceres, Pallas, Vesta over $\Delta t=\textcolor{black}{5}$ yr. The same initial conditions (J2000.0) have been used for both the integrations. The state vectors at the reference epoch have been retrieved from the NASA JPL Horizons system. The integrations have been performed in the  ICRF/J2000.0 reference frame, with the ecliptic and mean equinox of the reference epoch, centered at the Solar System Barycenter (SSB).} }\lb{EMB_Jupiter_CePaVe}
\end{figure}
The nominal peak-to-peak amplitude is of the order of \textcolor{black}{1000} m; given the $3-0.6\%$ level of uncertainty in the masses of such major asteroids\cite{CePaVe}, their mismodeled signature would amount to tens m.\\
  Such  effects would bias the general relativistic Schwarzschild  signal at a $\textcolor{black}{6\times 10^{-5}}$ level, being $\textcolor{black}{4}$ times larger than the Lense-Thirring one. Concerning the solar $J_2^{\odot}$, the \textcolor{black}{mismodelled action of} the major asteroids would represent up to $\textcolor{black}{30}\%$ of its signature. However, also in this case, the temporal pattern is different from the other ones.
\subsection{The Trans-Neptunian Objects}
Figure \ref{EMB_Jupiter_tnoring} is dedicated to the TNOs.
\begin{figure}[pb]
\centerline{\psfig{file=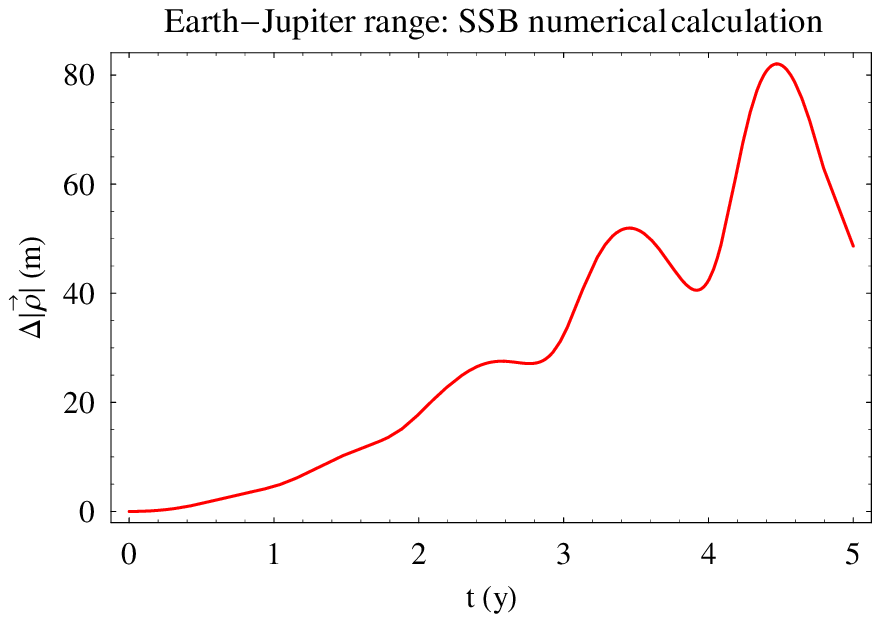
 ,width=4.7cm}}\vspace*{8pt}\caption{\textcolor{black}{Difference $\Delta |\vec{\rho}|\doteq |\vec{\rho}_{\rm P}|-|\vec{\rho}_{\rm R}|$ in the numerically integrated EMB-Jupiter ranges with and without the nominal perturbation due to the ring of Trans-Neptunian Objects with\protect\cite{Pit} $m_{\rm ring}=5.26\times 10^{-8}$M$_{\odot}$ and $R_{\rm ring}=43$ a.u. over $\Delta t=\textcolor{black}{5}$ yr. The same initial conditions (J2000.0) have been used for both the integrations. The state vectors at the reference epoch have been retrieved from the NASA JPL Horizons system. The integrations have been performed in the  ICRF/J2000.0 reference frame, with the ecliptic and mean equinox of the reference epoch, centered at the Solar System Barycenter (SSB).} }\lb{EMB_Jupiter_tnoring}
\end{figure}
It shows a signature with a nominal amplitude of $\textcolor{black}{80}$ m exhibiting a steadily increasing pattern. In regard to their corrupting action on the signals of interest, by assuming a $100\%$ uncertainty they would bias the Schwarzschild and $J_2$ effects at a $\textcolor{black}{1.6}\times 10^{-4}$ and $\textcolor{black}{80}\%$ level, respectively.
\subsection{Violation of the Strong Equivalence Principle}
In Figure \ref{EMB_Jupiter_eta} the effect of a SEP violation as large as $\eta=10^{-5}$  on the Earth-Jupiter range is plotted.
\begin{figure}[pb]
\centerline{\psfig{file=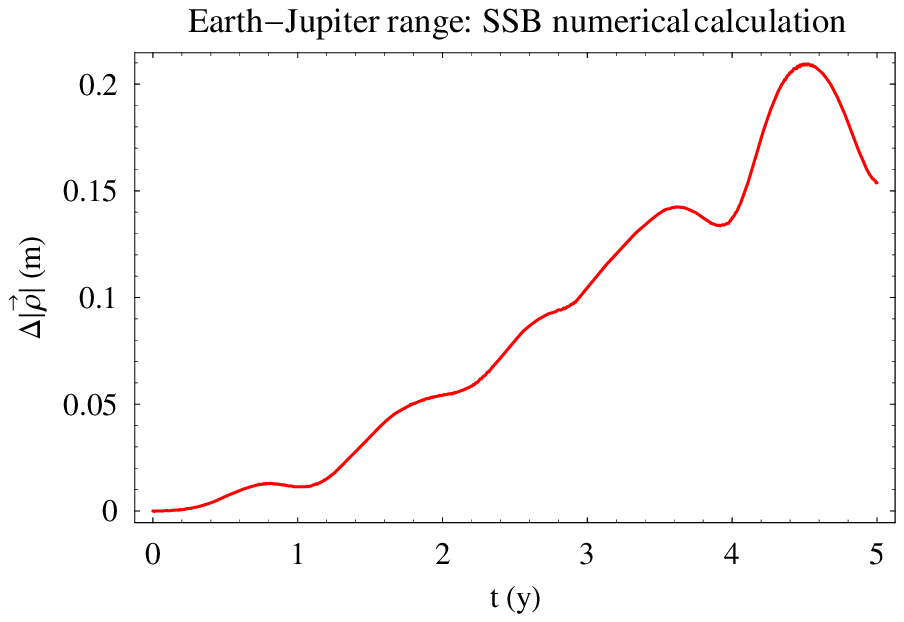
 ,width=4.7cm}}\vspace*{8pt}\caption{\textcolor{black}{Difference $\Delta |\vec{\rho}|\doteq |\vec{\rho}_{\rm P}|-|\vec{\rho}_{\rm R}|$ in the numerically integrated EMB-Jupiter ranges with and without the nominal perturbation due to a violation of SEP according to $\eta=10^{-5}$ over $\Delta t=\textcolor{black}{5}$ yr. The same initial conditions (J2000.0) have been used for both the integrations. The state vectors at the reference epoch have been retrieved from the NASA JPL Horizons system. The integrations have been performed in the  ICRF/J2000.0 reference frame, with the ecliptic and mean equinox of the reference epoch, centered at the Solar System Barycenter (SSB).} }\lb{EMB_Jupiter_eta}
\end{figure}
Its peak-to-peak amplitude is $\textcolor{black}{0.2}$ m, far too small to be detected in any foreseeable future also because it would be swamped by the competing Einsteinian and Newtonian signals.
\subsection{Secular variation of the Newtonian constant of gravitation}
The effect of a time variation of the Newtonian constant of gravitation as large as\cite{Pit} $\dot G/G=-5.9\times 10^{-14}$ yr$^{-1}$   is plotted  in Figure \ref{EMB_Jupiter_Gdot}.
\begin{figure}[pb]
\centerline{\psfig{file=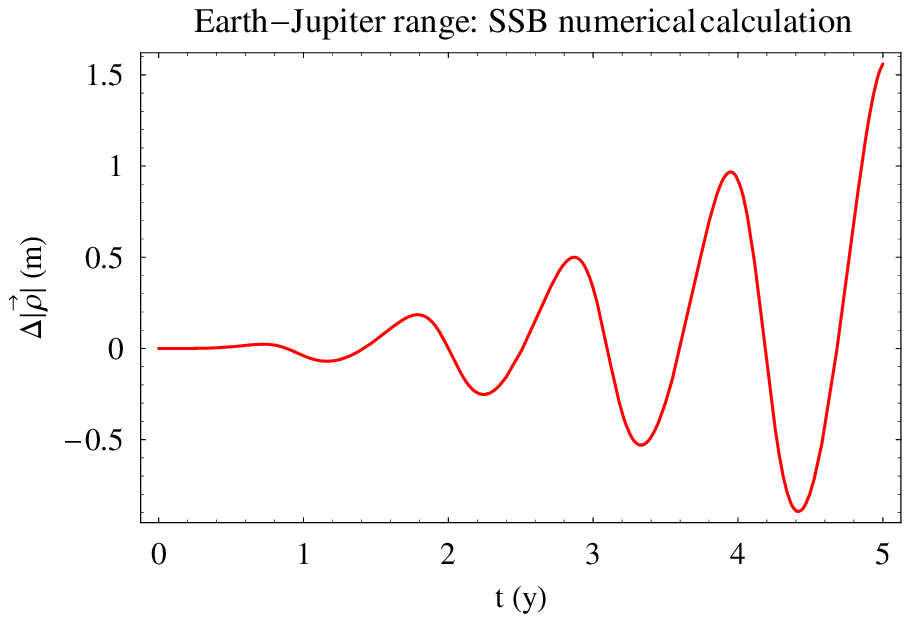
 ,width=4.7cm}}\vspace*{8pt}\caption{\textcolor{black}{Difference $\Delta |\vec{\rho}|\doteq |\vec{\rho}_{\rm P}|-|\vec{\rho}_{\rm R}|$ in the numerically integrated EMB-Jupiter ranges with and without the nominal perturbation due to a secular variation of $G$ as large as\protect\cite{Pit} $\dot G/G=-5.9\times 10^{-14}$ yr$^{-1}$ over $\Delta t=\textcolor{black}{5}$ yr. The same initial conditions (J2000.0) have been used for both the integrations. The state vectors at the reference epoch have been retrieved from the NASA JPL Horizons system. The integrations have been performed in the  ICRF/J2000.0 reference frame, with the ecliptic and mean equinox of the reference epoch, centered at the Solar System Barycenter (SSB).} }\lb{EMB_Jupiter_Gdot}
\end{figure}
Its nominal peak-to-peak amplitude is about $\textcolor{black}{2-2.5}$ m; also in this case, it is unlikely that any reasonable improvements in the ranging to Jupiter may allow for a detection of such a putative tiny effect. Moreover, it would be largely overwhelmed by the other relativistic and classical dynamical signals.
\subsection{The Pioneer Anomaly}
Figure \ref{EMB_Jupiter_Pio} shows the indirect effect of the standard form of the putative Pioneer anomaly on the jovian range through its direct action on Uranus, Neptune and Pluto.
\begin{figure}[pb]
\centerline{\psfig{file=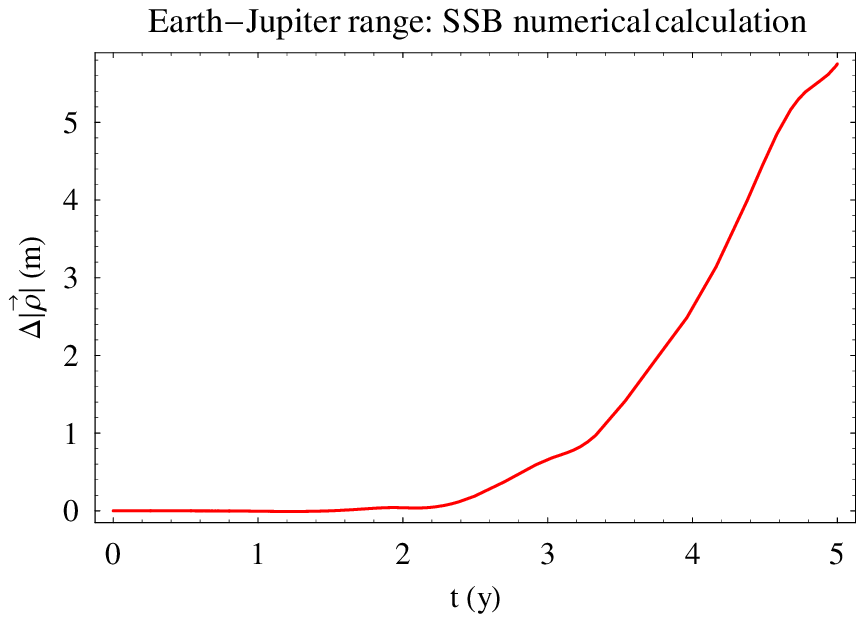
 ,width=4.7cm}}\vspace*{8pt}\caption{\textcolor{black}{Difference $\Delta |\vec{\rho}|\doteq |\vec{\rho}_{\rm P}|-|\vec{\rho}_{\rm R}|$ in the numerically integrated EMB-Jupiter ranges with and without the nominal perturbation due to a Pioneer-like constant and unform radial acceleration of\protect\cite{Pio} $A_{\rm Pio}=8.74\times 10^{-10}$ m s$^{-2}$ acting upon Uranus, Neptune, Pluto, Eris over $\Delta t=\textcolor{black}{5}$ yr. The same initial conditions (J2000.0) have been used for both the integrations. The state vectors at the reference epoch have been retrieved from the NASA JPL Horizons system. The integrations have been performed in the  ICRF/J2000.0 reference frame, with the ecliptic and mean equinox of the reference epoch, centered at the Solar System Barycenter (SSB).} }\lb{EMB_Jupiter_Pio}
\end{figure}
Its nominal effect is as large as $\textcolor{black}{5}$ m, likely undetectable. Also in this case, the other standard  dynamical effects would be  larger.
\subsection{Planet X}
The impact of a putative, distant  planet X on the range of Jupiter is shown in Figure \ref{EMB_Jupiter_X}-Figure \ref{EMB_Jupiter_X2}.
\begin{figure}[pb]
\centerline{\psfig{file=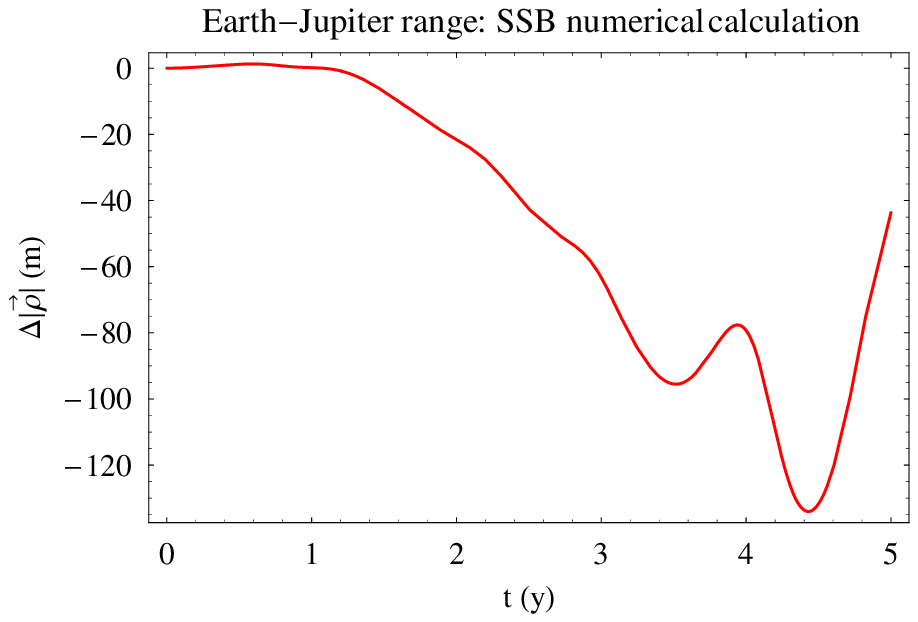
 ,width=4.7cm}}\vspace*{8pt}\caption{\textcolor{black}{Difference $\Delta |\vec{\rho}|\doteq |\vec{\rho}_{\rm P}|-|\vec{\rho}_{\rm R}|$ in the numerically integrated EMB-Jupiter ranges with and without the perturbation due to hypothetical remote planet X lying almost in the ecliptic with minimum tidal parameter\protect\cite{Iorio} ${\mathcal{K}}_{\rm X}=1.5\times 10^{-26}$ s$^{-2}$  over $\Delta t=\textcolor{black}{5}$ yr. The same initial conditions (J2000.0) have been used for both the integrations. The state vectors at the reference epoch have been retrieved from the NASA JPL Horizons system. The integrations have been performed in the  ICRF/J2000.0 reference frame, with the ecliptic and mean equinox of the reference epoch, centered at the Solar System Barycenter (SSB).} }\lb{EMB_Jupiter_X}
\end{figure}
\begin{figure}[pb]
\centerline{\psfig{file=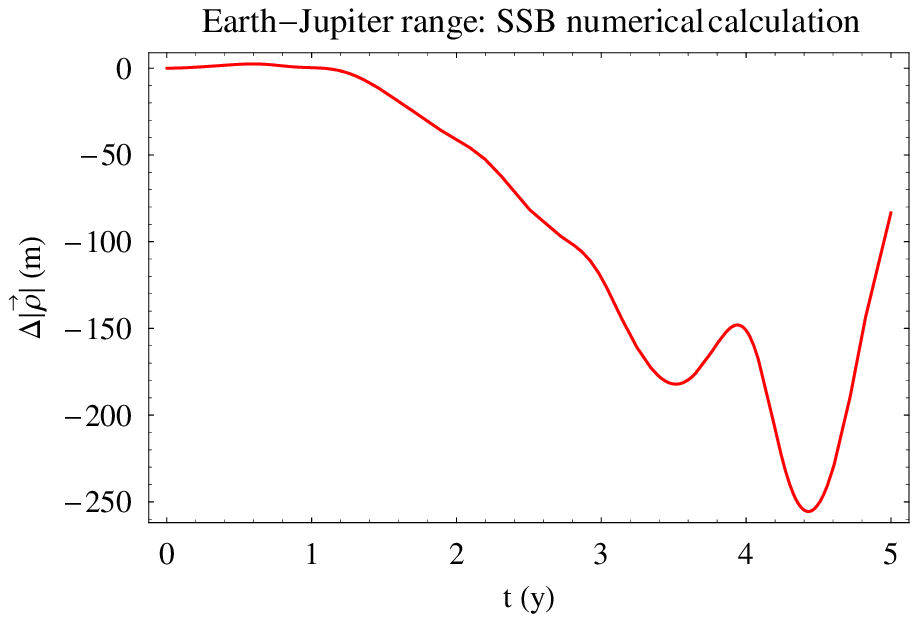
 ,width=4.7cm}}\vspace*{8pt}\caption{\textcolor{black}{Difference $\Delta |\vec{\rho}|\doteq |\vec{\rho}_{\rm P}|-|\vec{\rho}_{\rm R}|$ in the numerically integrated EMB-Jupiter ranges with and without the perturbation due to hypothetical remote planet X lying almost in the ecliptic with maximum tidal parameter\protect\cite{Iorio} ${\mathcal{K}}_{\rm X}=2.7\times 10^{-26}$ s$^{-2}$  over $\Delta t=\textcolor{black}{5}$ yr. The same initial conditions (J2000.0) have been used for both the integrations. The state vectors at the reference epoch have been retrieved from the NASA JPL Horizons system. The integrations have been performed in the  ICRF/J2000.0 reference frame, with the ecliptic and mean equinox of the reference epoch, centered at the Solar System Barycenter (SSB).} }\lb{EMB_Jupiter_X2}
\end{figure}
They refer to the minimum and maximum values of the tidal parameter $\mathcal{K}_{\rm X}$ of X as derived from the anomalous perihelion precession of Saturn\cite{Iorio} of $-6\pm 2$ mas cty$^{-1}$ preliminarily obtained from an initial data processing of the Cassini data themselves analyzed with the EPM ephemerides. The peak-to-peak amplitudes of the signals of X amount to $\textcolor{black}{120-250}$ m.
Such a putative effect is \textcolor{black}{roughly} comparable to the (nominal) signatures induced by the Sun's quadrupole,  the minor asteroids and the TNOs. However, the signature of X is different from those of such competing effects, apart from the ring of minor asteroids (Figure \ref{EMB_Jupiter_astring}).
}
\section{Earth-Saturn range}\lb{saturn}
After the Cassini spacecraft started its \virg{grand tour} of the Saturnian system, it has been possible to drastically increase the accuracy of the orbit determination of the ringed planet through direct ranging to Cassini itself. Figure B-20 of Ref.~\refcite{DE421} shows the range residuals of Saturn from 2004 to 2006 constructed with the DE421 ephemerides from Cassini normal points; processing of extended data records of Cassini is currently ongoing, \textcolor{black}{so that we will consider an integration time span of 5 yr}. The range residuals of Figure B-20 in Ref.~\refcite{DE421} are accurate at 10 m level. Also a pair of range residuals from close encounters with Voyager 1 (1980) and Voyager 2 (1982) are shown: they are almost one order of magnitude less accurate.
\subsection{The Schwarzschild field of the Sun}
Figure \ref{EMB_Saturn_Schwa} shows the Schwarzschild perturbation of the  range of Saturn over $\Delta t=\textcolor{black}{5}$ yr. Its peak-to-peak amplitude amounts to \textcolor{black}{$5\times 10^{5}$} m.
\begin{figure}[pb]
\centerline{\psfig{file=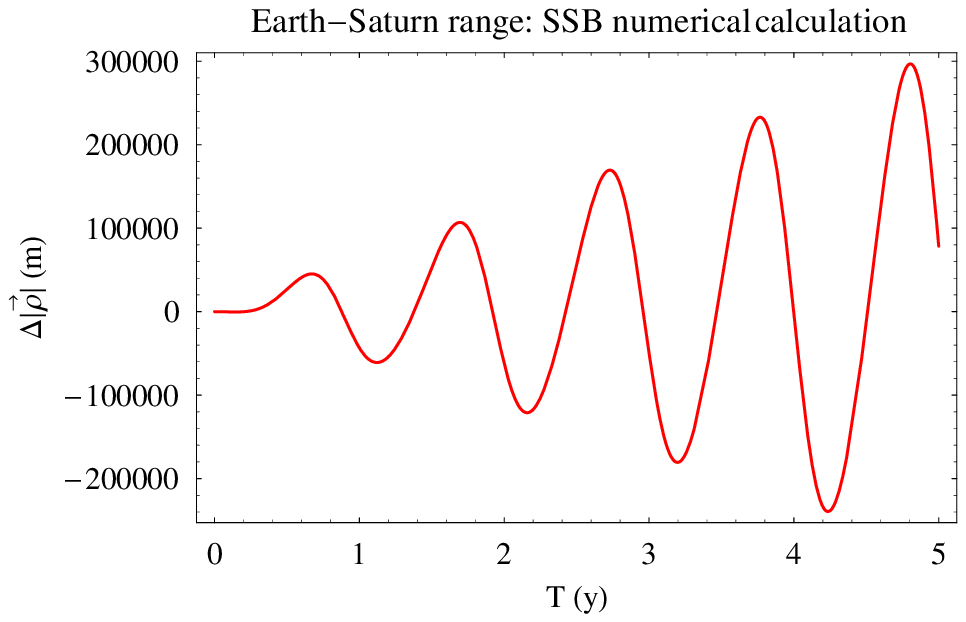
 ,width=4.7cm}}\vspace*{8pt}\caption{Difference $\Delta |\vec{\rho}|\doteq |\vec{\rho}_{\rm P}|-|\vec{\rho}_{\rm R}|$ in the numerically integrated EMB-Saturn ranges with and without the perturbation due to the Sun's Schwarzschild  field over $\Delta t=\textcolor{black}{5}$ yr. The same initial conditions (J2000.0) have been used for both the integrations. The state vectors at the reference epoch have been retrieved from the NASA JPL Horizons system. The integrations have been performed in the  ICRF/J2000.0 reference frame, with the ecliptic and mean equinox of the reference epoch, centered at the Solar System Barycenter (SSB). }\lb{EMB_Saturn_Schwa}
\end{figure}
This implies a measurement accuracy of \textcolor{black}{$2\times 10^{-5}$}, given the present-day level of uncertainty in the Cassini ranging residuals.
\subsection{The  oblateness of the Sun}
The Sun's oblateness effect on the Saturn range, computed in a solar equatorial frame,  is depicted in Figure \ref{EMB_Saturn_J2}.
\begin{figure}[pb]
\centerline{\psfig{file=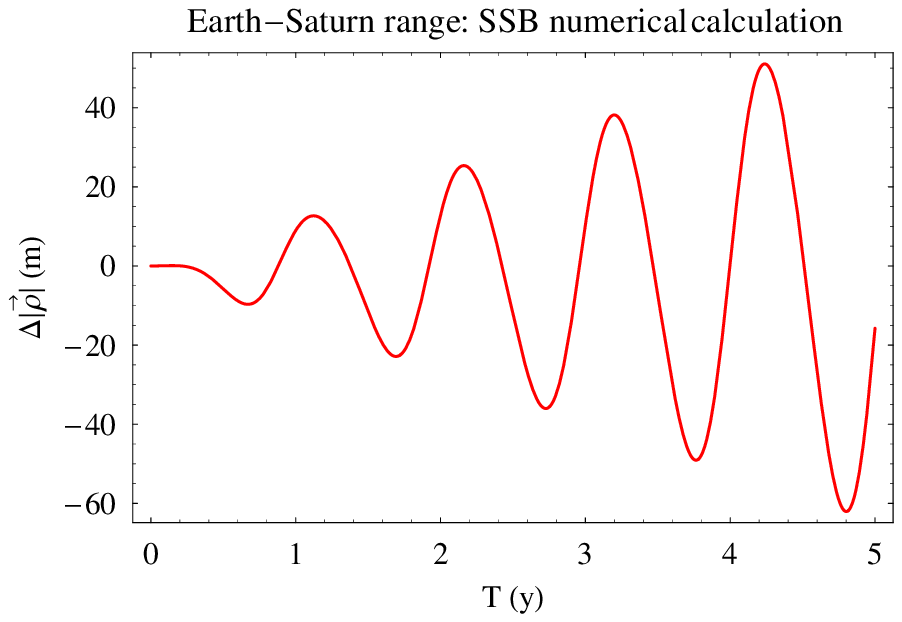
 ,width=4.7cm}}\vspace*{8pt}\caption{Difference $\Delta |\vec{\rho}|\doteq |\vec{\rho}_{\rm P}|-|\vec{\rho}_{\rm R}|$ in the numerically integrated EMB-Saturn ranges with and without the nominal perturbation due to the Sun's quadrupole mass moment $J_2=2.0\times 10^{-7}$ over $\Delta t=\textcolor{black}{5}$ yr. The same initial conditions (J2000.0) have been used for both the integrations. The state vectors at the reference epoch have been retrieved from the NASA JPL Horizons system. The integrations have been performed in the  ICRF/J2000.0 reference frame, with the mean equinox of the reference epoch and the reference $\{xy\}$ plane rotated from the mean ecliptic of the epoch to the Sun's equator, centered at the Solar System Barycenter (SSB). }\lb{EMB_Saturn_J2}
\end{figure}
Its nominal peak-to-peak amplitude is about \textcolor{black}{100} m. Thus, the relative accuracy in measuring it is \textcolor{black}{somewhat} modest, amounting to just \textcolor{black}{$1\times 10^{-1}$}. A $10\%$ uncertainty in the solar $J_2$ implies a mismodeled signal of \textcolor{black}{10} m. It represents a bias of $2\times 10^{-5}$ on the Schwarzschild signature, which, however, has a different pattern.
\subsection{The Lense-Thirring effect of the Sun}
The Lense-Thirring range perturbation for Saturn, computed in a frame with the $z$ axis aligned with the Sun's spin axis,  is illustrated in Figure \ref{EMB_Saturn_LT}.
\begin{figure}[pb]
\centerline{\psfig{file=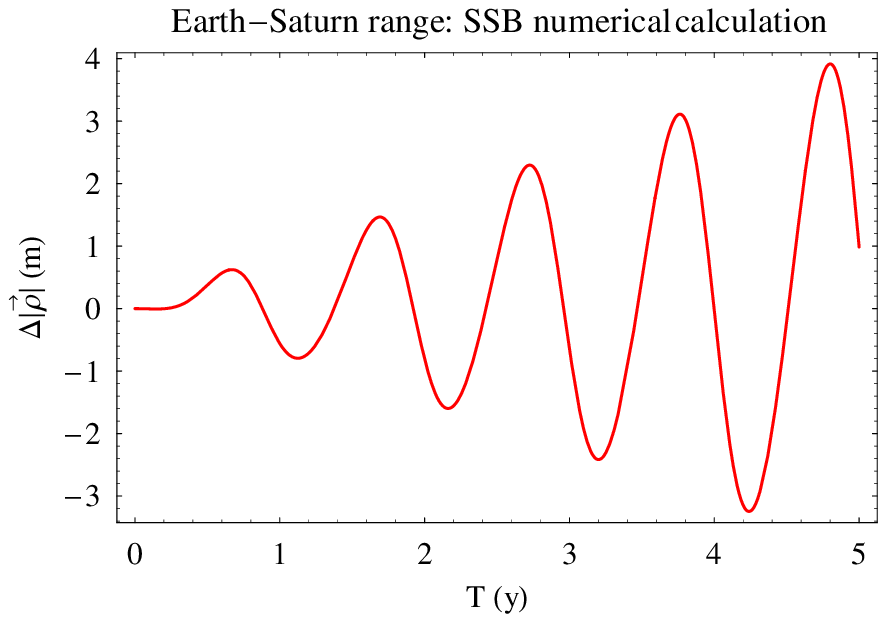
 ,width=4.7cm}}\vspace*{8pt}\caption{Difference $\Delta |\vec{\rho}|\doteq |\vec{\rho}_{\rm P}|-|\vec{\rho}_{\rm R}|$ in the numerically integrated EMB-Saturn ranges with and without the perturbation due to the Sun's Lense-Thirring field over $\Delta t=\textcolor{black}{5}$ yr. The same initial conditions (J2000.0) have been used for both the integrations. The state vectors at the reference epoch have been retrieved from the NASA JPL Horizons system. The integrations have been performed in the  ICRF/J2000.0 reference frame, with the mean equinox of the reference epoch and the reference $\{xy\}$ plane rotated from the mean ecliptic of the epoch to the Sun's equator, centered at the Solar System Barycenter (SSB). }\lb{EMB_Saturn_LT}
\end{figure}
Its peak-to-peak amplitude is approximately \textcolor{black}{7} m, too small to be detected with the present-day ranging accuracy to Cassini. It would affect the Schwarzschild perturbation at $1\times 10^{-5}$ level, while the $J_2$ effect would be biased at $7\times 10^{-2}$ level. The time signature of the gravitomagnetic shift is similar to the larger relativistic effect, but it is different from that due to the Sun's oblateness.
\subsection{The ring of the minor asteroids and Ceres, Pallas and Vesta}
The nominal effect of the ring of the minor asteroids on the Saturnian range is in Figure \ref{EMB_Saturn_astring}.
\begin{figure}[pb]
\centerline{\psfig{file=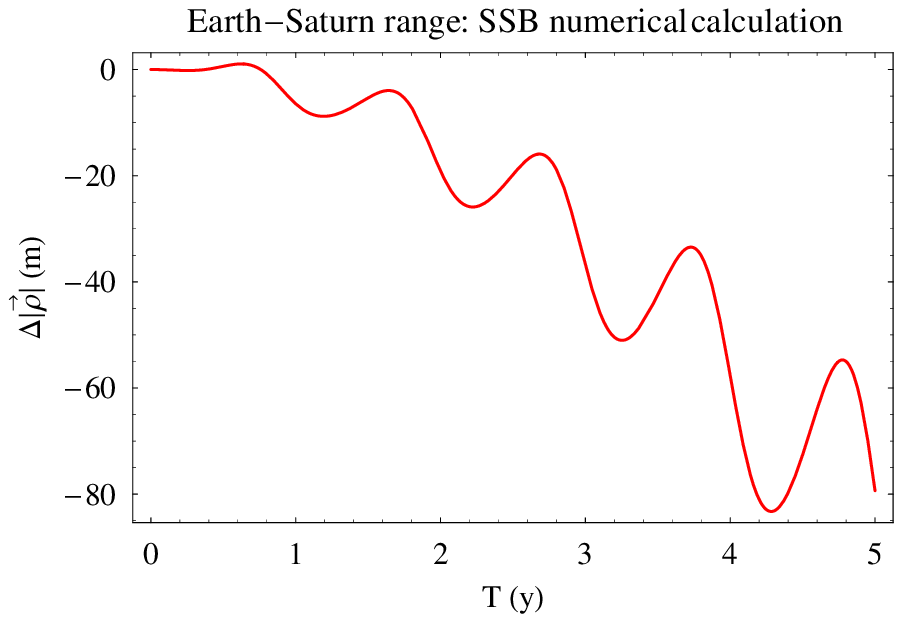
 ,width=4.7cm}}\vspace*{8pt}\caption{Difference $\Delta |\vec{\rho}|\doteq |\vec{\rho}_{\rm P}|-|\vec{\rho}_{\rm R}|$ in the numerically integrated EMB-Saturn ranges with and without the nominal perturbation due to the ring of minor asteroids with\protect\cite{Fienga} $m_{\rm ring}=1\times 10^{-10}$M$_{\odot}$  and $R_{\rm ring}=3.14$ a.u. over $\Delta t=\textcolor{black}{5}$ yr. The same initial conditions (J2000.0) have been used for both the integrations. The state vectors at the reference epoch have been retrieved from the NASA JPL Horizons system. The integrations have been performed in the  ICRF/J2000.0 reference frame, with the ecliptic and mean equinox of the reference epoch, centered at the Solar System Barycenter (SSB). }\lb{EMB_Saturn_astring}
\end{figure}
Its peak-to-peak amplitude is as large as \textcolor{black}{80} m, barely detectable with the current Cassini ranging. The corresponding mismodeled signal would amount to \textcolor{black}{24} m, compatible with the \textcolor{black}{currently available} Cassini range residuals \textcolor{black}{spanning 2 yr}. Such a source of systematic alias impacts the Schwarzschild perturbation at a \textcolor{black}{$5\times 10^{-5}$} level, while the $J_2$ signal is biased by it at a  $2\times 10^{-1}$ level. The Lense-Thirring signal is overwhelmed by the alias due to the minor asteroids.
The nominal perturbation on the Saturn range by the combined action of Ceres, Pallas, Vesta is
computed illustrated in Figure \ref{EMB_Saturn_CePaVe}.
\begin{figure}[pb]
\centerline{\psfig{file=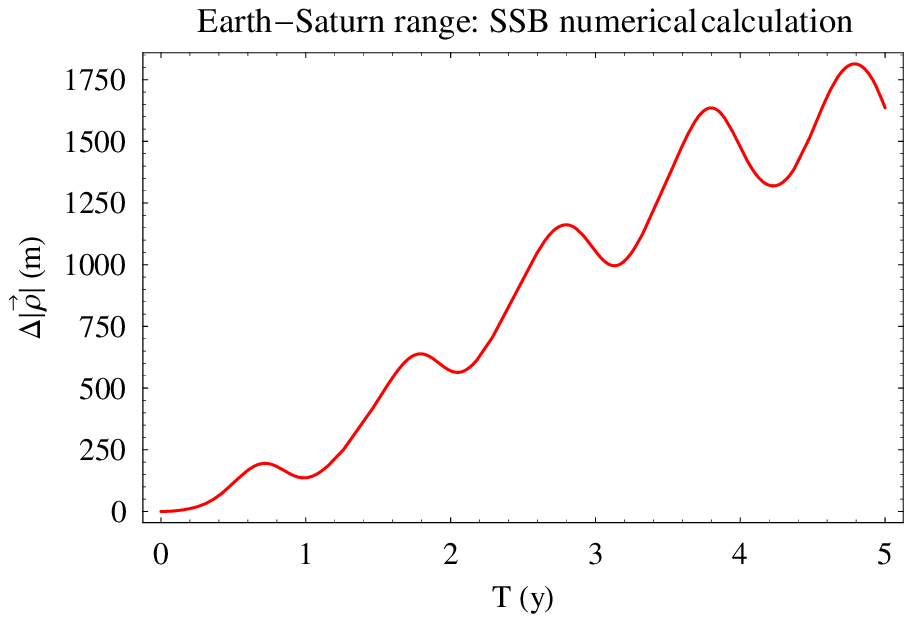
 ,width=4.7cm}}\vspace*{8pt}\caption{Difference $\Delta |\vec{\rho}|\doteq |\vec{\rho}_{\rm P}|-|\vec{\rho}_{\rm R}|$ in the numerically integrated EMB-Saturn ranges with and without the nominal perturbation due to\protect\cite{CePaVe} Ceres, Pallas, Vesta  over $\Delta t=\textcolor{black}{5}$ yr. The same initial conditions (J2000.0) have been used for both the integrations. The state vectors at the reference epoch have been retrieved from the NASA JPL Horizons system. The integrations have been performed in the  ICRF/J2000.0 reference frame, with the ecliptic and mean equinox of the reference epoch, centered at the Solar System Barycenter (SSB). }\lb{EMB_Saturn_CePaVe}
\end{figure}
The  peak-to-peak amplitude amounts to \textcolor{black}{1750} m, measurable at about \textcolor{black}{$6\times 10^{-3}$} according to the current level of accuracy of the ranging to Cassini. On the other hand, the peak-to-peak amplitude of the mismodeled signature of the three asteroids, whose dynamical action is included in the dynamical models of all the modern ephemerides,  is just 6 m. It would overwhelm the Lense-Thirring effect, while the relative bias  on the Schwarzschild and $J_2$ range perturbations would be $4\times 10^{-5},2\times 10^{-1}$, respectively.
\subsection{The Trans-Neptunian Objects}
The action of the TNOs on the Saturn range is more effective. Figure \ref{EMB_Saturn_tnoring} shows that it amounts to \textcolor{black}{200} m.
\begin{figure}[pb]
\centerline{\psfig{file=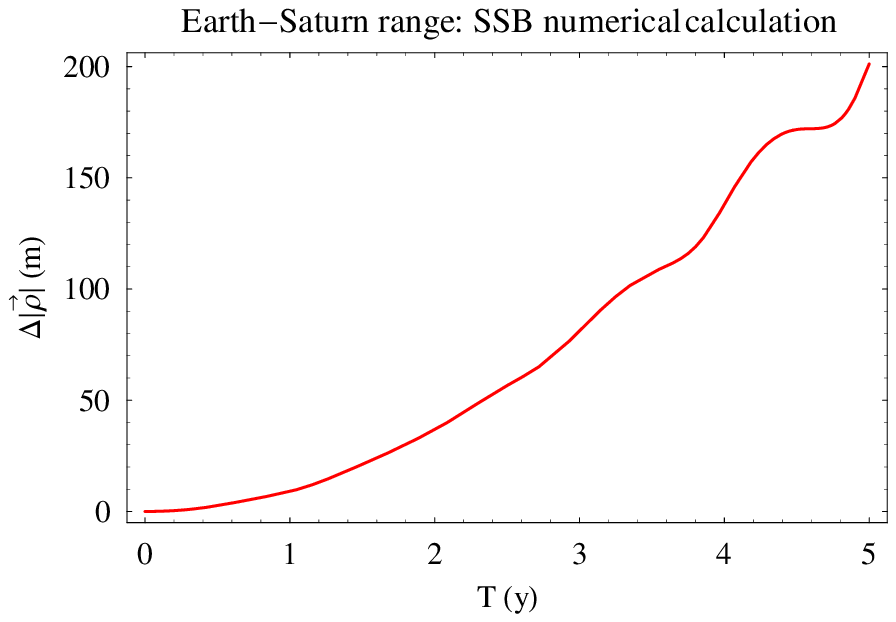
 ,width=4.7cm}}\vspace*{8pt}\caption{Difference $\Delta |\vec{\rho}|\doteq |\vec{\rho}_{\rm P}|-|\vec{\rho}_{\rm R}|$ in the numerically integrated EMB-Saturn ranges with and without the nominal perturbation due to the ring of Trans-Neptunian Objects with\protect\cite{Pit}  $m_{\rm ring}=5.26\times 10^{-8}$M$_{\odot}$ and $R_{\rm ring}=43$ a.u. over $\Delta t=\textcolor{black}{5}$ yr. The same initial conditions (J2000.0) have been used for both the integrations. The state vectors at the reference epoch have been retrieved from the NASA JPL Horizons system. The integrations have been performed in the  ICRF/J2000.0 reference frame, with the ecliptic and mean equinox of the reference epoch, centered at the Solar System Barycenter (SSB). }\lb{EMB_Saturn_tnoring}
\end{figure}
In principle, it would be measurable by the Cassini ranging. The absence  of such a signal  in the \textcolor{black}{present-day} range residuals may be due to the fact that the dynamical action
of the TNOs was not modeled in the DE421 ephemerides used to produce them, so that it is likely that part of the TNOs signature, \textcolor{black}{amounting to 50 m over 2 yr}, has been removed due to the fitting of the initial conditions. The Schwarzschild effect is aliased by them at a \textcolor{black}{$4\times 10^{-4}$} level, while the $J_2$ and the Lense-Thirring signals would be swamped by the TNOs. It is interesting to note that their temporal evolution is different from that of the minor asteroids.
\subsection{Violation of the Strong Equivalence Principle}
Figure \ref{EMB_Saturn_eta} shows the effect of a violation of SEP for $\eta=10^{-5}$ on the Saturn range.
\begin{figure}[pb]
\centerline{\psfig{file=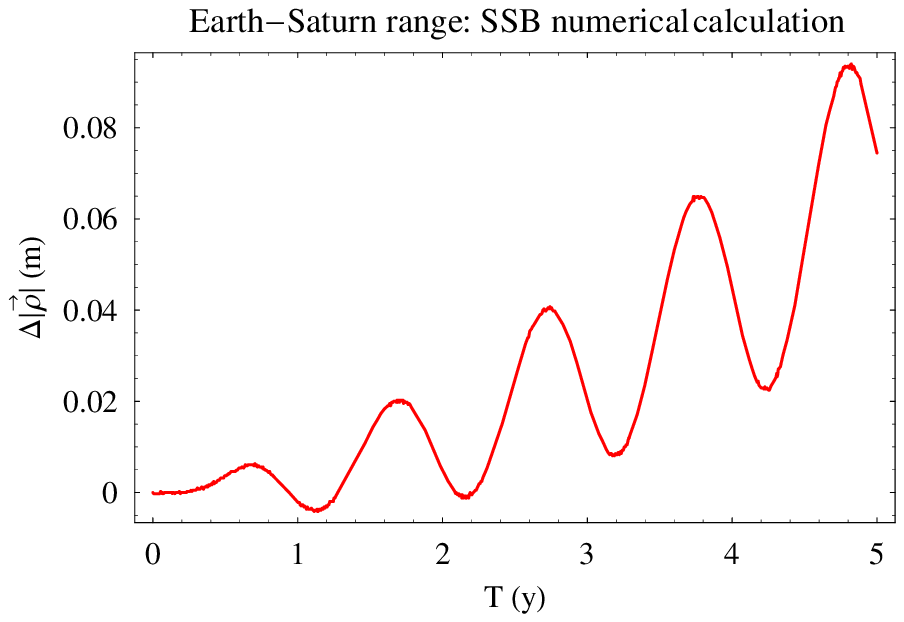
 ,width=4.7cm}}\vspace*{8pt}\caption{Difference $\Delta |\vec{\rho}|\doteq |\vec{\rho}_{\rm P}|-|\vec{\rho}_{\rm R}|$ in the numerically integrated EMB-Saturn ranges with and without the nominal perturbation due to a violation of SEP according to $\eta=10^{-5}$ over $\Delta t=\textcolor{black}{5}$ yr. The same initial conditions (J2000.0) have been used for both the integrations. The state vectors at the reference epoch have been retrieved from the NASA JPL Horizons system. The integrations have been performed in the  ICRF/J2000.0 reference frame, with the ecliptic and mean equinox of the reference epoch, centered at the Solar System Barycenter (SSB). }\lb{EMB_Saturn_eta}
\end{figure}
Also in this case it is quite negligible because it is as large as \textcolor{black}{9} cm, almost two orders of magnitude smaller that the present-day level of accuracy in the Cassini ranging. Moreover, also the other dynamical effects considered would completely swamp such a signal.
\subsection{Secular variation of the Newtonian constant of gravitation}
A secular variation of the Newtonian constant of gravitation as large as that reported by Pitjeva in  Ref.~\refcite{Pit} would produce as signal as in Figure  \ref{EMB_Saturn_Gdot}.
\begin{figure}[pb]
\centerline{\psfig{file=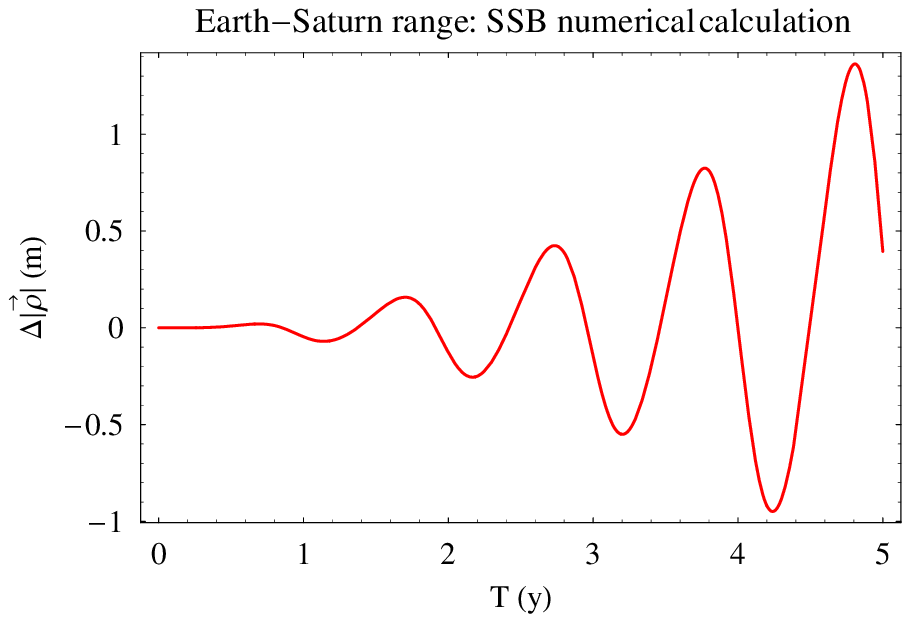
 ,width=4.7cm}}\vspace*{8pt}\caption{Difference $\Delta |\vec{\rho}|\doteq |\vec{\rho}_{\rm P}|-|\vec{\rho}_{\rm R}|$ in the numerically integrated EMB-Saturn ranges with and without the nominal perturbation due to a secular variation of $G$ as large as\protect\cite{Pit} $\dot G/G=-5.9\times 10^{-14}$ yr$^{-1}$  over $\Delta t=\textcolor{black}{5}$ yr. The same initial conditions (J2000.0) have been used for both the integrations. The state vectors at the reference epoch have been retrieved from the NASA JPL Horizons system. The integrations have been performed in the  ICRF/J2000.0 reference frame, with the ecliptic and mean equinox of the reference epoch, centered at the Solar System Barycenter (SSB). }\lb{EMB_Saturn_Gdot}
\end{figure}
Its peak-to-peak amplitude is \textcolor{black}{2.5} m, too small to be detected with the present-day ranging to Cassini. Also all the other standard Newtonian and relativistic effects, modeled or not, would be orders of magnitude larger; for example, it must be recalled that the mismodeled action of Ceres, Pallas, Vesta is as large as \textcolor{black}{17} m.
\subsection{The Pioneer Anomaly}
The indirect effect of the Pioneer anomaly on the range of Saturn is depicted in Figure \ref{EMB_Saturn_Pio}.
\begin{figure}[pb]
\centerline{\psfig{file=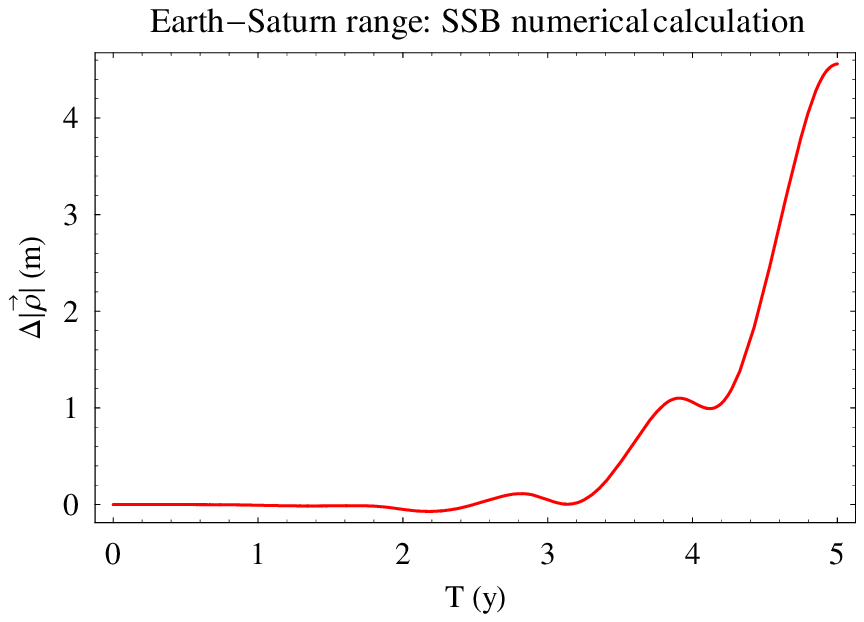
 ,width=4.7cm}}\vspace*{8pt}\caption{Difference $\Delta |\vec{\rho}|\doteq |\vec{\rho}_{\rm P}|-|\vec{\rho}_{\rm R}|$ in the numerically integrated EMB-Saturn ranges with and without the nominal perturbation due to a Pioneer-like constant and unform radial acceleration of\protect\cite{Pio} $A_{\rm Pio}=8.74\times 10^{-10}$ m s$^{-2}$ acting upon Uranus, Neptune, Pluto, Eris over $\Delta t=\textcolor{black}{5}$ yr. The same initial conditions (J2000.0) have been used for both the integrations. The state vectors at the reference epoch have been retrieved from the NASA JPL Horizons system. The integrations have been performed in the  ICRF/J2000.0 reference frame, with the ecliptic and mean equinox of the reference epoch, centered at the Solar System Barycenter (SSB). }\lb{EMB_Saturn_Pio}
\end{figure}
It would be undetectable because it would be as large as \textcolor{black}{4 m} over $\Delta t =$ \textcolor{black}{5} yr. Moreover, such a signal would be totally overwhelmed by the other dynamical effects considered so far, both of Newtonian and relativistic origin.
\subsection{Planet X}
In Figure \ref{EMB_Saturn_X}-Figure \ref{EMB_Saturn_X2} the impact of a putative, distant  planet X on the range of Saturn is depicted.
\begin{figure}[pb]
\centerline{\psfig{file=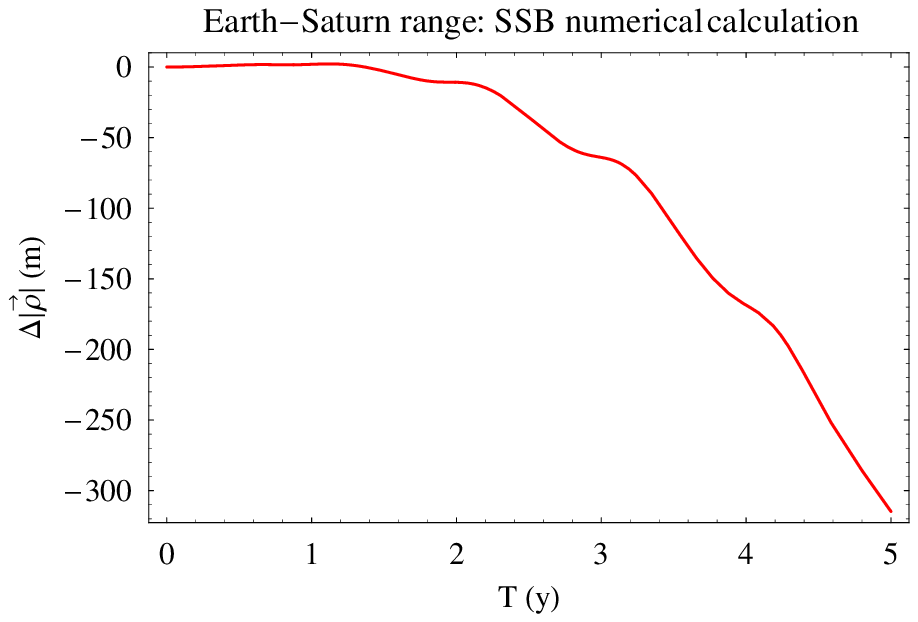
 ,width=4.7cm}}\vspace*{8pt}\caption{Difference $\Delta |\vec{\rho}|\doteq |\vec{\rho}_{\rm P}|-|\vec{\rho}_{\rm R}|$ in the numerically integrated EMB-Saturn ranges with and without the perturbation due to hypothetical remote planet X lying almost in the ecliptic with minimum tidal parameter\protect\cite{Iorio} ${\mathcal{K}}_{\rm X}=1.5\times 10^{-26}$ s$^{-2}$  over $\Delta t=\textcolor{black}{5}$ yr. The same initial conditions (J2000.0) have been used for both the integrations. The state vectors at the reference epoch have been retrieved from the NASA JPL Horizons system. The integrations have been performed in the  ICRF/J2000.0 reference frame, with the ecliptic and mean equinox of the reference epoch, centered at the Solar System Barycenter (SSB). }\lb{EMB_Saturn_X}
\end{figure}
\begin{figure}[pb]
\centerline{\psfig{file=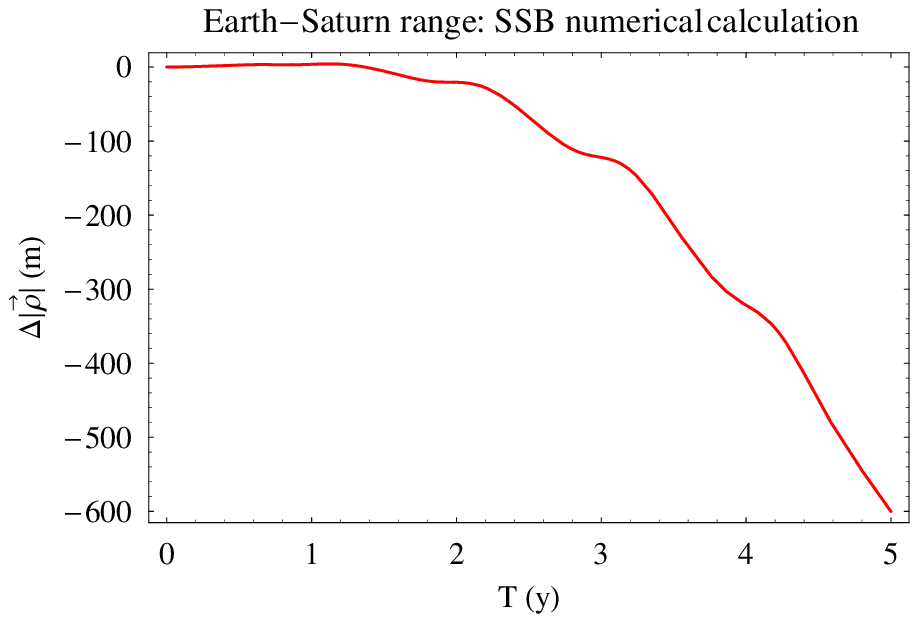
 ,width=4.7cm}}\vspace*{8pt}\caption{Difference $\Delta |\vec{\rho}|\doteq |\vec{\rho}_{\rm P}|-|\vec{\rho}_{\rm R}|$ in the numerically integrated EMB-Saturn ranges with and without the perturbation due to hypothetical remote planet X lying almost in the ecliptic with maximum tidal parameter\protect\cite{Iorio} ${\mathcal{K}}_{\rm X}=2.7\times 10^{-26}$ s$^{-2}$  over $\Delta t=\textcolor{black}{5}$ yr. The same initial conditions (J2000.0) have been used for both the integrations. The state vectors at the reference epoch have been retrieved from the NASA JPL Horizons system. The integrations have been performed in the  ICRF/J2000.0 reference frame, with the ecliptic and mean equinox of the reference epoch, centered at the Solar System Barycenter (SSB). }\lb{EMB_Saturn_X2}
\end{figure}
Let us recall that they refer to the minimum and maximum values of the tidal parameter $\mathcal{K}_{\rm X}$ of X as derived from the anomalous perihelion precession of Saturn of $-6\pm 2$ mas cty$^{-1}$ obtained from the Cassini data themselves analyzed with the EPM ephemerides. The peak-to-peak amplitudes of the signals of X \textcolor{black}{, over $\Delta t=5$ yr,} amount to \textcolor{black}{$300-600$} m. \textcolor{black}{If we consider just 2 yr, the magnitude of the X signatures} practically coincide with the present-day accuracy of the Cassini ranging. It is interesting to note that they are not in contrast with the Cassini range residuals obtained with the DE421 ephemerides of Figure B-20 of Ref.~\refcite{DE421}. The dynamical action of X was not included  in the force models of the DE421 ephemerides; as explained before, fitting the initial conditions may partially or totally remove an unmodeled signature.
The action of the TNOs would be a competitor of X, at least on such a timescales. Anyway, the patterns of the two signals are different. Instead, $J_2$, given the present-day level of uncertainty in it, would not alias the X signature. The Lense-Thirring effect is \textcolor{black}{two orders} of magnitude smaller.
\section{Summary and conclusions}\lb{conclu}
In view of a possible future implementation of some interplanetary laser ranging facilities accurate to cm-level,  we have numerically investigated how the ranges between the Earth and all the inner planets plus \textcolor{black}{ Jupiter and } Saturn are affected by certain Newtonian and non-Newtonian dynamical effects by simultaneously integrating the equations of motion of all the major bodies of the solar system plus some minor bodies of it (Ceres, Pallas, Vesta, Pluto, Eris) in the SSB reference frame over a time span \textcolor{black}{2} years long, \textcolor{black}{ apart from Mars, \textcolor{black}{Jupiter} and Saturn for which we adopted \textcolor{black}{5} years. }
One of the major goals of the forthcoming or planned interplanetary ranging missions like, e.g., BepiColombo is the accurate (of the order of, or better than $10^{-6}$) determination of the PPN parameters $\gamma$ and $\beta$ discriminating various metric theories of gravity. To this aim,
it must be recalled that the observable used in actual interplanetary ranging tests of post-Newtonian gravity consists of two \textcolor{black}{gravitationally affected} parts: the one, purely relativistic, connected with the Shapiro delay of the propagation of the electromagnetic waves  induced by the Schwarzschild field of the Sun, and the other one due to the reciprocal Earth-planet/spacecraft orbital dynamics; reaching exquisite accuracies in only measuring the Shapiro delay  is useless if the orbital component is known less accurately. That is why
we paid attention to several Newtonian perturbations on the planetary ranges which may be viewed as sources of systematic uncertainty in the main general relativistic Schwarzschild signals of interest. The same hold\textcolor{black}{s} for other Newtonian and non-Newtonian target effects as well.

It turns out that the general relativistic gravitomagnetic Lense-Thirring effect of the Sun, not modeled so far either in the planetary ephemerides or in the analyses of some spacecraft-based future missions like, e.g., BepiColombo, does actually fall within the measurability domain of future  cm-level ranging devices. The more favorable situation occurs for Mercury because the relative measurement accuracy is of the order of $2-5\times 10^{-3}$ by assuming a $4.5-10$ cm uncertainty in the Earth-Mercury ranging, as expected for BepiColombo over some years of operations. It is $2-5\times 10^{-2}$ for Venus and \textcolor{black}{$1.2-2.5\times 10^{-2}$} for Mars by assuming the same level of uncertainty in the corresponding planetary ranging over \textcolor{black}{5 yr}. In the case of Saturn, the peak-to-peak amplitude of the gravitomagnetic range signal is \textcolor{black}{7} m \textcolor{black}{over 5 yr}, too small by about a factor \textcolor{black}{1.5}  with respect to the present-day level of accuracy in the ranging to Cassini. If not properly modeled and solved-for, the Lense-Thirring effect may also impact the determination of other Newtonian and post-Newtonian parameters at a nonnegligible level, given the high accuracy with which their measurement is pursued. For example, in the case of BepiColombo the expected accuracy in determining  $\gamma$ and $\beta$ from the range perturbation due to the  Schwarzschild field of the Sun is of the order of $10^{-6}$; the Lense-Thirring range perturbation would impact the Schwarzschild one at $4\times 10^{-5}$ level. Another goal of the BepiColombo mission is a  measurement of the Sun's quadrupole mass moment $J_2$ accurate to $10^{-2}$; the unmodeled Lense-Thirring effect would bias it at  $10^{-1}$ level. From the point of view of a measurement of the Sun's gravitomagnetic field itself, it results that
a major concern would be the solar oblateness; it should be known at a $10^{-2}$ level of accuracy-which is just the goal of BepiColombo-to allow for a  reduction of its aliasing impact on  the Lense-Thirring signal down to just $17\%$.
The ring of the minor asteroids should be taken into account as well because its mismodeling would impact the gravitomagnetic signal at about $7\times 10^{-2}$. The lingering uncertainty in the masses of Ceres, Pallas, Vesta translates into a potential bias of about $3\times 10^{-2}$. The TNOs, not modeled so far apart from the EPM ephemerides, would nominally affect it at a $4.5\times 10^{-2}$ level; it must be considered that there is currently a high uncertainty in their mass. However, it must be noted that the patterns of such sources of systematic bias are different with respect to the gravitomagnetic one.
About Venus and Mars, the measurement of their Lense-Thirring range perturbations would be made difficult by the mismodeled signals due to $J_2$, the ring of the minor asteroids, Ceres, Pallas, Vesta and the TNOs because their magnitudes are often as large as, or even larger than, the gravitomagnetic ones, although their temporal signatures would be different.

A Newtonian dynamical effect that has been investigated as a potential source of systematic uncertainty in the planetary range signals of interest is the ring of the minor asteroids. Its nominal signatures  would be  detectable because they are  of the order of 4 m (Mercury), 3 m (Venus), \textcolor{black}{40} m (Mars), \textcolor{black}{\textcolor{black}{250} m (Jupiter)} \textcolor{black}{80} m (Saturn). However, \textcolor{black}{it} is currently  modeled in the present-day ephemerides, and the uncertainty in its mass is of the order of $30\%$. Thus, the peak-to-peak amplitudes of the mismodeled effects are 1.2 m (Mercury), 90 cm (Venus), \textcolor{black}{12} m (Mars), \textcolor{black}{\textcolor{black}{75} m (Jupiter)} and \textcolor{black}{24} m (Saturn). Concerning BepiColombo \textcolor{black}{and Mercury}, the impact of such aliasing signals on the Schwarzschild, $J_2$ and Lense-Thirring range perturbations is $3\times 10^{-6}, 4\times 10^{-3}, 7\times 10^{-2}$, respectively; it is not negligible with respect to the expected levels of accuracy. Moreover, it must be recalled that in some proposed spacecraft-based tests of post-Newtonian gravity the target accuracies in measuring $\gamma$ and $\beta$ may be as high as $10^{-8}-10^{-9}$.
In the case of Venus,  the uncertainties in the masses of Ceres, Pallas, Vesta translate into a relative bias on the Schwarzschild, $J_2$ and Lense-Thirring range perturbations of $4\times 10^{-7}, 1\times 10^{-3}, 3\times 10^{-2}$, respectively.
The situation for Mars, \textcolor{black}{which is an}other possible candidate for implementing accurate interplanetary ranging, is less favorable because the relative uncertainties in the three signals of interest are \textcolor{black}{$4.3\times 10^{-5}, 2\times 10^{-1}, 3$}, respectively.  Saturn should be considered as well in view of possible Cassini ranging tests of post-Newtonian gravity. The impact of the mismodeling in the  ring of the minor asteroids is \textcolor{black}{$5\times 10^{-5}$} for the Schwarzschild range perturbation. The dynamical action of the three major asteroids, i.e. Ceres, Pallas, Vesta, is currently modeled in all the modern planetary ephemerides, but the present-day $10^{-3}-10^{-2}$ level of uncertainty in their masses would induce mismodeled signatures which cannot be neglected with respect to the goal of accurate measurements of the Newtonian post-Newtonian parameters of interest. Indeed, in the case of Mercury the systematic uncertainties in the Schwarzschild, $J_2$ and Lense-Thirring range perturbations are $1\times 10^{-6}, 2\times 10^{-3}, 3\times 10^{-2}$, respectively. For Venus their impact on the first two effects is $8\times 10^{-6}, 2\times 10^{-2}$, respectively, while the gravitomagnetic signature would be overwhelmed. \textcolor{black}{It is worse} with Mars, for which the alias in the Schwarzschild and $J_2$ range perturbations due to the mismodeling in the three large asteroids is \textcolor{black}{$5\times 10^{-5}, 2\times 10^{-1}$}, respectively. \textcolor{black}{However, it must be recalled that more asteroids do actually perturb the orbit of Mars at a nonnegligible level.} The gravitoelectric signal in the range of Saturn would be affected at a \textcolor{black}{$3.5\times 10^{-5}$} level by them.

A classical dynamical effect not considered so far in some ephemerides and mission analyses is the action of the TNOs, which we modeled as a massive ring. It turns out that it may impact  some high-precision tests of Newtonian and post-Newtonian gravity just at the level of desired accuracy. In the case of BepiColombo, the bias of the TNOs on the Schwarzschild and $J_2$ range signals amounts to $2\times 10^{-6}$ and $3\times 10^{-3}$, respectively. The TNOs' impact on the gravitoelectric and $J_2$ range of Venus is as large as $4\times 10^{-6}, 1.2\times 10^{-2}$, respectively. More effective is their action on the Mars range. Indeed, in this case their bias on the Schwarzschild and $J_2$ ranges is \textcolor{black}{$2\times 10^{-5}$} and \textcolor{black}{$8\times 10^{-2}$}, respectively. It maybe interesting to note that the TNOs impact the Schwarzschild range signal of Saturn at \textcolor{black}{$4\times 10^{-4}$} level; it should be taken into account in possible, future Cassini ranging-based tests of post-Newtonian gravity at the ringed planet. \textcolor{black}{A similar level of bias by the TNOs on the gravitoelectric range signal occurs also for Jupiter.}

We also examined other non-Newtonian dynamical effects like a SEP violation through the $\eta$ parameter, a secular variation of the Newtonian constant of gravitation and the indirect effect of the Pioneer anomaly on the inner planets through the altered action of the giant planets, putatively acted upon by it, on them. Concerning the SEP violation, since $\eta$ is currently known at $10^{-4}$ level from LLR, we looked at the case $\eta=10^{-5}$. It turns out that the largest effect occurs for Mars and Saturn amounting to \textcolor{black}{$5-9$} cm \textcolor{black}{($\Delta t=5$ yr)}; for Mercury and Venus the corresponding SEP signals are as large as 6 mm and 8 mm, respectively, \textcolor{black}{over 2 yr}. They are realistically too small to be detectable; moreover, they would be completely overwhelmed by all the other unmodeled/mismodeled standard Newtonian and relativistic effects. Recently, a statistically significative secular decrease of $G$ of the order of $10^{-14}$ yr$^{-1}$ has been preliminarily reported by E.V. Pitjeva; thus, we looked at interplanetary range signals due to such an effect as well. It turns out that BepiColombo would, perhaps, be able to detect the corresponding range perturbation for Mercury of 60 cm (peak-to-peak maximum amplitude) with a relative accuracy of $1-2\times 10^{-1}$; it would impact the Schwarzschild, $J_2$ and Lense-Thirring signals at $1\times 10^{-6}, 2\times 10^{-3}, 3\times 10^{-2}$, respectively. Conversely, the Lense-Thirring and TNOs signals, if not modeled, would severely affect the putative $\dot G/G$ signature; the same also holds for the mismodeled signature of Ceres, Pallas, Vesta. For Venus \textcolor{black}{($\Delta t=2$ yr)} and Mars \textcolor{black}{($\Delta t=5$ yr)} the range signals would be as large as \textcolor{black}{$0.07-1$} m, likely too small to be detectable, also because of the huge biasing action of the other competing standard Newtonian and relativistic effects. The  $\dot G/G$ range signal for Saturn is  \textcolor{black}{2 m over 5 yr}, which is beyond the current capabilities of the ranging to Cassini. The indirect effects of the Pioneer anomaly on the interplanetary ranges examined are far too small amounting to a few mm for \textcolor{black}{Mercury and Venus ($\Delta t=2$ yr)} (for \textcolor{black}{Jupiter and} \textcolor{black}{Saturn} it is up to \textcolor{black}{4\textcolor{black}{-5} m over 5 yr}).

Finally, we looked at the perturbations on the interplanetary range signals that a distant, planetary-sized body X may induce. We considered for its tidal parameter the minimum and maximum values obtained by analyzing a putative anomalous precession of the perihelion of Saturn recently determined by Pitjeva and Fienga et al. from preliminary analyses of some years of Cassini normal points with the latest versions of the EPM and INPOP ephemerides. Future analyses of extended data sets of Cassini should shed more light on the genuine existence of such a phenomenon. Our investigations show that the range signals caused by X would be well measurable with future cm-level ranging devices. Indeed, the peak-to-peak amplitudes of its signatures for Mercury and Venus  are $1.5-3$ m and $3-5$ m, respectively. For \textcolor{black}{Mars, \textcolor{black}{Jupiter} and} Saturn the peak-to-peak amplitude\textcolor{black}{s} of the X range perturbation\textcolor{black}{s are $10-25$ m, \textcolor{black}{$120-250$ m} and} \textcolor{black}{$300-600$} m\textcolor{black}{, respectively, over $\Delta t=5$ yr.}
\textcolor{black}{The \textcolor{black}{figures for Saturn} are quite larger} than  the present-day level of accuracy in the ranging to Cassini.

Table \ref{TAVOLA} summarizes our results.

Our analysis can also be  extended to other putative, nonconventional gravitational accelerations induced by modified models of gravity. Moreover, it may be helpful in analyzing also other scenarios involving ranging to spacecraft not necessarily orbiting a given planet or satellite. Indeed, the ASTROD project involves the use of three spacecraft ranging coherently with one another using lasers: one should be located near the Earth at one of the Lagrange points L1/L2, while the other two should move along separate solar orbits. The ASTROD I concept relies upon laser ranging from laser stations on the Earth to one spacecraft in solar orbit. LATOR is based on
the use of a laser transceiver terminal on the International
Space Station (ISS) and two spacecraft placed in $\sim 1$ a.u. heliocentric orbits. For such missions the expected level of accuracy in determining $\beta$ and $\gamma$ is of the order of $10^{-8}-10^{-9}$ after some years of operations.

\textcolor{black}{At the end, let us elucidate certain limitations of the present study and trace some possible routes for further investigations.
Our analysis  pretends to be neither complete nor definitive because, for example, it only accounts for the actions of the three major asteroids. This is particularly true for Mars whose orbit is actually perturbed by a much larger number of such small bodies; it is well known that modern ephemerides, built up with  larger human, material and computational resources than those available for the present study,  include up to 300 biggest asteroids acting in a nonnegligible way on the red planet. Another issue necessarily left out by our analysis is a complete investigation of the effective detectability of the dynamical effects considered. To this aim, it would have been necessary to implement numerical integrations fitting the initial conditions to the real observations as well since in the actual data processing such a procedure may absorb, to an extent to be quantitatively determined, the signature one is interested in.
In this framework, also the impact of the noise in the observations on the measurability of the different dynamical effects should be addressed, along with their mutual separability through a covariance analysis.
However, such  important tasks, which are outside the scopes of the present analysis, may be the goal for further work by skilful independent teams of astronomers routinely engaged in producing even more and more accurate ephemerides. }
\begin{table}[ph]
\tbl{
Maximum peak-to-peak nominal amplitudes, in m, of the Earth-planet range signals  due to the dynamical effects listed in the left column for Mercury, Venus, Mars, \textcolor{black}{Jupiter} and Saturn. \textcolor{black}{The time span used for Mercury and Venus  is $\Delta t=2$ yr, while for Mars, \textcolor{black}{Jupiter} and Saturn it is $\Delta t=5$ yr.} We adopted the standard value\protect\cite{Fienga} $J_2=2.0\times 10^{-7}$  for the quadrupole mass moment of the Sun, while for its angular  momentum we used\protect\cite{Pij1,Pij2} $S=190.0\times 10^{39}$ kg m$^2$ s$^{-1}$  from helioseismology. For the ring of the minor asteroids we used\protect\cite{Fienga}  $m_{\rm ring}=1\times 10^{-10}{\rm M}_{\odot},R_{\rm ring}=3.14$ a.u., while for the TNOs, modeled as massive ring as well, we adopted\protect\cite{Pit} $m_{\rm ring}=5.26\times 10^{-8}{\rm M}_{\odot}, R_{\rm ring}=43$ a.u. The masses of the major asteroids Ceres Pallas, Vesta have been retrieved from Ref.~\protect\refcite{CePaVe}. The magnitude of the SEP violation investigated is $\eta=10^{-5}$. The secular variation of $G$ as been accounted for according to Ref.~\protect\refcite{Pit} $\dot G/G = -5.9\times 10^{-14}$ yr$^{-1}$. The Pioneer anomaly has been included in the forces acting on the outer planets with its standard magnitude of\protect\cite{Pio} $8.74\times 10^{-10}$ m s$^{-2}$, while for the tidal parameter of X we used\protect\cite{Iorio} $GM_{\rm X}/r_{\rm X}^3 =(2.1\pm 0.6)\times 10^{-26}$ s$^{-2}$, obtained from the perihelion precession of Saturn\protect\cite{Fienga,Pitjou}.
\label{TAVOLA}
}
{
\begin{tabular}{@{}cccccc@{}}
\toprule
Dynamical effect & Mercury & Venus & Mars & \textcolor{black}{Jupiter} & Saturn \\
\colrule
Solar Schwarzschild & $4\times 10^5$ & $1.2\times 10^5$ & \textcolor{black}{$2.5\times 10^5$} & \textcolor{black}{$5\times 10^5$} & \textcolor{black}{$5\times 10^5$}\\
Solar $J_2$ & 300 & 40 & \textcolor{black}{70} & \textcolor{black}{110} & \textcolor{black}{100}\\
Solar Lense-Thirring & 17.5 & 2 & \textcolor{black}{4} & \textcolor{black}{7} & \textcolor{black}{7} \\
Ring of minor asteroids & 4 & 3 & \textcolor{black}{40} & \textcolor{black}{250} & \textcolor{black}{80}\\
Ceres, Pallas, Vesta & 80 & 175 & \textcolor{black}{1400} & \textcolor{black}{1000} & \textcolor{black}{1750} \\
TNOs & 0.8 & 0.5 & \textcolor{black}{5} & \textcolor{black}{80} & \textcolor{black}{200} \\
SEP  & $6\times 10^{-3}$ & $8\times 10^{-3}$ & \textcolor{black}{0.05} & \textcolor{black}{$0.2$} & \textcolor{black}{$0.09$} \\
$\dot G/G$ & $0.6$ & 0.07 & \textcolor{black}{1}  & \textcolor{black}{$2$} & \textcolor{black}{2}\\
Pioneer anomaly & $4\times 10^{-3}$ & $5\times 10^{-3}$ & \textcolor{black}{$0.3$}  & \textcolor{black}{$5$} & \textcolor{black}{$4$} \\
Planet X & $1.5-3$ & $3-5$ & $\textcolor{black}{10-25}$ & \textcolor{black}{$120-250$} & $\textcolor{black}{300-600}$\\
\botrule
\end{tabular}
}
\end{table}

\end{document}